\documentclass[twocolumn]{aastex701}
\usepackage{listings}
\usepackage{styles}

\shorttitle{alabi}
\shortauthors{birky et al.}
\graphicspath{{./}{figures/}}

\begin{document}

\title{\sc \texttt{ALABI:} Active Learning for Accelerated Bayesian Inference}

\correspondingauthor{Jessica Birky}
\email{jessbirky@proton.me}

\author[0000-0002-7961-6881]{Jessica Birky}
\email{jbirky@uw.edu}
\affiliation{Department of Astronomy, University of Washington, 3910 15th Avenue NE, Seattle, WA 98195, USA}
\affiliation{DiRAC Institute, University of Washington, 3910 15th Avenue NE, Seattle, WA 98195, USA} 

\author[0000-0001-6487-5445]{Rory Barnes}
\email{rkb9@uw.edu}
\affiliation{Department of Astronomy, University of Washington, 3910 15th Avenue NE, Seattle, WA 98195, USA}

\begin{abstract}

We present Active Learning for Accelerated Bayesian Inference (\alabi): an open-source Python package for performing Bayesian inference with computationally expensive models. Given a forward model and observational data to construct a likelihood and priors, \alabi\ uses a Gaussian Process (GP) surrogate model trained to predict posterior probability as a function of input parameters, and employs active learning to iteratively improve GP predictive performance in high-likelihood regions where the GP is most uncertain. \alabi\ provides a uniform interface for using Markov chain Monte Carlo (MCMC) with different packages, including the affine-invariant sampler \emcee, and nested samplers \dynesty, \multinest, and \ultranest. This approach facilitates accurate estimation of the desired posterior distribution, while reducing the number of computationally expensive model evaluations required by factors of thousands. We demonstrate the performance of \alabi\ on a variety of test cases, including where inference is challenging due to complex posterior structure or high dimensionality. We show that \alabi\ offers a substantial improvement for likelihood functions with evaluation times $\gtrsim 1$\,s, speeding up MCMC computations by a factor of $10-1000\times$ when tested on problems with up to 64 dimensions.  \\

\end{abstract}


\section{Introduction} \label{sec:intro}

The exponential growth in data availability and model complexity across computational sciences is transforming the landscape of scientific inference. 
Although this data abundance offers unprecedented opportunities for discovery, it also presents computational challenges that strain traditional analytical approaches.
Simulations, or forward models, that predict the outcomes of different conditions are highly useful for gaining insight from datasets.
Bayesian inference provides a natural framework for incorporating prior knowledge, quantifying uncertainty, and making robust decisions under uncertainty \citep{gelman_bayesian_2013,murphy_machine_2012}. However, the computational demands of Bayesian methods often render them impractical for complex, high-dimensional problems that characterize modern scientific applications.

The computational bottleneck in Bayesian inference stems primarily from the need to evaluate complex likelihood functions to properly explore the parameter space. Traditional approaches, particularly Markov chain Monte Carlo (MCMC) methods \citep{hogg_data_2010,speagle_conceptual_2020}, require extensive sampling to achieve convergence, making them prohibitively expensive for models with costly likelihood evaluations. This limitation is especially pronounced in applications that involve large-scale simulations, partial differential equations, or models with intricate hierarchical structures, where each likelihood evaluation may require substantial computational resources.

Recent advances in machine learning offer promising avenues for addressing these computational challenges \citep[e.g.,][]{cranmer_frontier_2020}. This paper introduces \alabi\footnote{\href{https://github.com/jbirky/alabi}{https://github.com/jbirky/alabi}}: a comprehensive framework for accelerated Bayesian inference that uses machine learning techniques to overcome the computational limitations of traditional methods. Our approach addresses the critical need for scalable inference tools in computationally expensive scientific models by providing efficient posterior approximations while maintaining statistical rigor. We present both the theoretical foundations and the practical implementation of \alabi, demonstrating its effectiveness in various benchmarks and providing guidance to practitioners facing similar computational challenges.

Other similar packages for performing active learning with Gaussian processes include \approxposterior\ \citep{fleming_approxposterior_2018}, \code{botorch} \citep{balandat2020botorch}, \code{BayesianOptimization} \citep{nogueira_2014}, \code{scikit-optimize} \citep{head_scikit-optimizescikit-optimize_2020}, \code{GPry} \cite{El_Gammal_2023}, \code{BOBE} \cite{cohen2026}, and \code{MCMC-BO} \cite{yi_2024}. One common challenge between these types of framework is that they require many steps to train accurately. These steps often involve choosing different algorithms and tuning many free parameters, which can be a steep learning curve for new users and require many trials and experiments to understand. Furthermore, there are a number of common failure modes in which the steps of the process can easily become numerically unstable and fail (such as poorly conditioned matrices, poor hyperparameter optimization, poor GP convergence, or poor MCMC convergence). In practice, these challenges make it difficult to train surrogate models for Bayesian problems, as it is often difficult to diagnose where an underlying problem is occurring and identify the setting that will fix the issue.

If not conditioned well, GP models also suffer from the ``curse of dimensionality'' problem, in which the number of training points needed to achieve convergence scales exponentially with the number of dimensions. This effect is well documented in the GP literature (e.g., see \citealt{binois_survey_2022} for a detailed review). In this paper, we discuss how \alabi\ improves its training performance for high dimensions over a na\"ively implemented GP.

The goal of this work is not to present novel algorithms or replacements for existing codes, but rather to develop a ``best practice'' guide on how a user might choose different configurations for their problem. \alabi\ is designed to create a flexible, yet user-friendly framework for easily testing between different GP surrogate models and/or MCMC samplers, and includes comprehensive documentation on a variety of benchmark problems.

The remainder of this paper is structured as follows. Section \ref{sec:methods} presents the theoretical framework underlying \alabi, including the machine learning techniques employed for posterior approximation. Section \ref{sec:implement} details the software implementation, user interface, and provides comprehensive examples in Python. Section \ref{sec:results} evaluates \alabi's performance across multiple benchmark problems, comparing accuracy and computational efficiency against traditional methods. Finally, Section \ref{sec:discussion} discusses the practical implications of our results, identifies current limitations, and provides guidance for selecting appropriate inference strategies for different problem cases. \\

\section{Methods} \label{sec:methods}

The \alabi\ Python package provides a user-friendly and efficient platform for performing Bayesian inference with computationally expensive models. Given a likelihood function defined by the user (usually constructed from a forward model and observational data), \alabi\ trains a Gaussian process (GP) surrogate model to emulate the likelihood function. As described in this section, the GP is trained in an iterative process designed to minimize the overall computational expense. This training process consists of two parts: the Gaussian process and an active learning algorithm.

\subsection{Bayesian Inference} \label{subsec:bayes}

In typical Bayesian inference problems, the goal is usually to sample the posterior distribution:
\begin{equation}
    P(\Theta|\mathcal{D}) = \frac{P(\mathcal{D}|\Theta) \cdot P(\Theta)}{P(\mathcal{D})},
\end{equation}
where $P(\mathcal{D}|\Theta)$ is the likelihood probability of a model with parameters $\Theta$ given data $\mathcal{D}$. $P(\Theta)$ is known as the prior which incorporates previously known information, $P(\mathcal{D})$ is known as the evidence, and $P(\Theta|\mathcal{D})$ is the posterior.

The fundamental challenge in computing the posterior distribution lies in evaluating the evidence term $P(\mathrm{\mathcal{D}})$, which requires integrating over the entire parameter space: 
\begin{equation}
    P(\mathcal{D}) = \int P(\mathcal{D}|\Theta) \cdot P(\Theta) \, d\Theta.
\end{equation}
This integral is analytically intractable for most realistic models, particularly for high-dimensional parameter spaces. Direct numerical integration methods become computationally prohibitive because the volume of the parameter space grows exponentially with each additional dimension. MCMC methods circumvent this problem by generating samples from the posterior distribution without explicitly computing the evidence term, since the normalization constant cancels out in the acceptance ratios used by MCMC algorithms. However, MCMC sampling remains computationally intensive because it requires evaluating the likelihood function $P(\mathcal{D}|\Theta)$ at each proposed step in the chain, and achieving adequate sampling of the posterior often requires millions of likelihood evaluations to ensure proper convergence, particularly when dealing with complex, multi-modal, or highly correlated parameter spaces.

\begin{figure}[htbp]
    \centering
    \adjustbox{scale=0.85,center}{\includegraphics{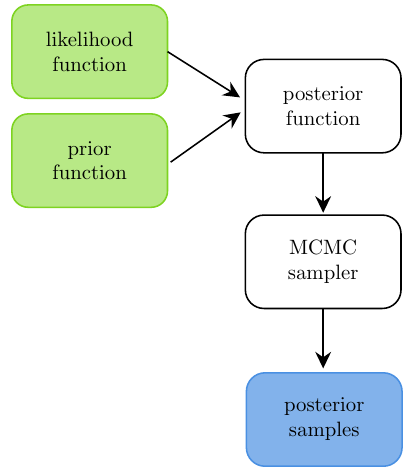}}
    \caption{Schematic showing the steps involved in using a traditional MCMC sampler.}
    \label{fig:diagram1}
\end{figure}

\begin{figure}[htbp]
    \centering
    \adjustbox{scale=0.9,center}{\includegraphics{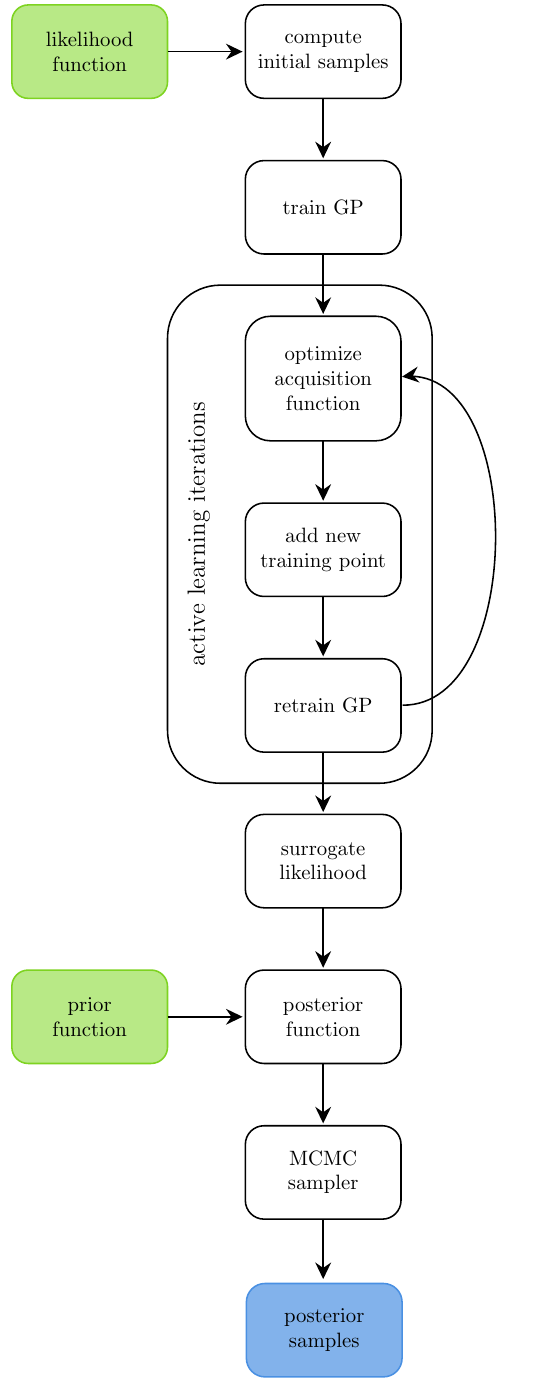}}
    \caption{Schematic of the \alabi\ framework. The user defines their likelihood and prior functions as inputs to \alabi, and \alabi\ returns samples of the posterior.}
    \label{fig:diagram2}
\end{figure}

For likelihood functions that are slow to evaluate, \alabi\ is designed to speed up MCMC computations by replacing the likelihood function with a \emph{surrogate model}, which once trained, can quickly be evaluated for millions of samples for robust MCMC sampling.
As a conceptual overview, Figure \ref{fig:diagram1} shows the workflow for traditional MCMC inference, while Figure \ref{fig:diagram2} shows the workflow for MCMC inference using \alabi. In both cases the user input and output are the same: the user defines the likelihood function and prior function for their problem, and the framework returns the computed posterior samples. While under the hood \alabi\ requires more steps, \alabi\ is modularized such that it remains easy to use on any black-box likelihood function. This modularity allows the user a high degree of flexibility to define their likelihood and prior functions as is appropriate for their problem, while abstracting away the details of surrogate model training.

\subsection{Gaussian Process}

The key difference between \alabi\ (Fig. \ref{fig:diagram2}) and standard MCMC (Fig. \ref{fig:diagram1}) is the use of GP regression to emulate the likelihood function.
GPs provide a flexible and powerful framework for modeling and inference in various domains, including machine learning, statistics, and computational science. A GP defines a distribution over functions, such that any finite set of function values follows a multivariate Gaussian distribution. This property makes GPs particularly suitable for modeling complex and nonlinear relationships in data. 
In this section we give a brief overview of GPs; for an in-depth review see \cite{rasmussen_gaussian_2006}. 

A GP is defined by a mean function, denoted as $m(\mathbf{x})$, and covariance function (or kernel), denoted as $k(\mathbf{x}, \mathbf{x}')$. The mean function captures the overall trend of the underlying function, while the covariance function characterizes the dependency structure and smoothness properties. Mathematically, a GP is defined by a multivariate Gaussian:
\begin{equation}
    \mathcal{GP}(\mathbf{x}) \sim \mathcal{N}(\mu(\mathbf{x}), k(\mathbf{x}, \mathbf{x}')),
\end{equation}
where $\mathbf{x} \in \mathbb{R}^m$ is an input vector with $m$ dimensions,  $\mu(\mathbf{x})$ is the mean function representing the expected value of the function at each input point x, and $k(\mathbf{x}, \mathbf{x}')$ is the covariance function modeling the correlation between function values at different input points $\mathbf{x}$ and $\mathbf{x}'$. This kernel function, chosen by the user, effectively quantifies the smoothness and spatial properties of the GP. 

An example of a commonly used kernel function that performs well on a wide range of functions is the squared exponential kernel:
\begin{equation} 
    k_{\rm SE}(r) = A \cdot \exp \left( - \frac{1}{2} r^2 \right),
    \label{eqn:exp_squared_kernel}
\end{equation}
where $r$ is the scaled separation ($r = |\mathbf{x} - \mathbf{x}'| / \ell$)
between two given input points $\mathbf{x}$ and $\mathbf{x}'$, $\mathbf{\ell}$ is a kernel hyperparameter that defines the length scale of the correlation in each dimension, and $A$ is the hyperparameter that defines the amplitude. 

Other options for kernels compatible with \alabi include the Matern-3/2 kernel:
\begin{equation}
    k_{\rm M 3/2}(r) = A \cdot  \left(1 + \sqrt{3r^2} \right) \exp(-\sqrt{3r^2}),
    \label{eqn:matern32_kernel}
\end{equation}
or the Matern-5/2 kernel:
\begin{equation}
    k_{\rm M5/2}(r) = A \cdot  \left(1 + \sqrt{5r^2} + \frac{5r^2}{3} \right) \exp(-\sqrt{5r^2}).
    \label{eqn:matern52_kernel}
\end{equation}

\begin{figure}
    \centering
    \includegraphics[width=0.5\textwidth]{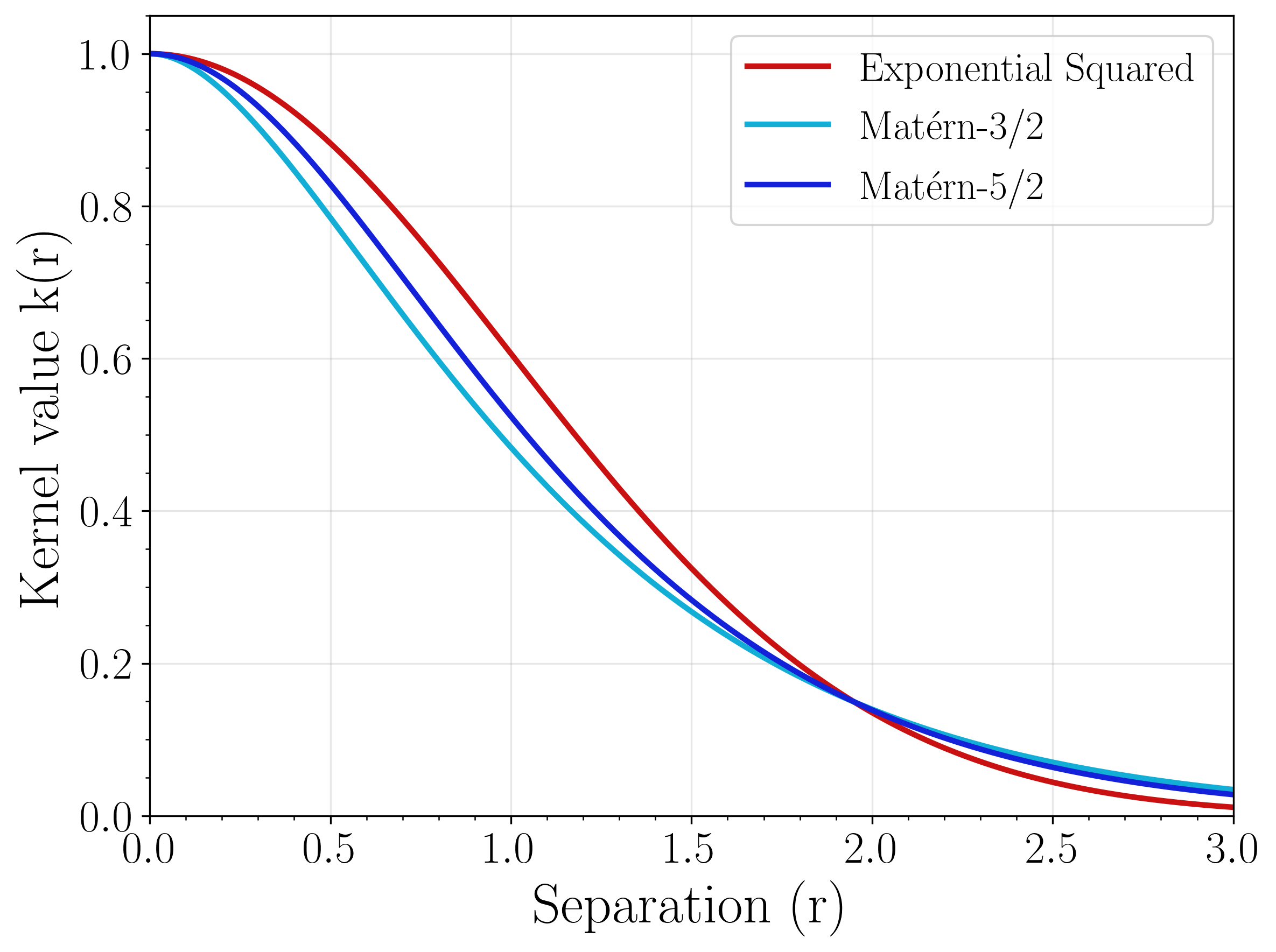}
    \caption{Plot comparing the shape of different kernel functions, with length scale $\ell=1$ and amplitude $A=1$.}
    \label{fig:kernel_comparison}
\end{figure}

Figure \ref{fig:kernel_comparison} compares the exponential squared, Matern-3/2, and Matern-5/2 kernels. While the shapes of each function are qualitatively similar, the Matern-3/2 kernel has a steeper peak at small separations, while the exponential squared kernel is flatter at small separations. 
In Section \ref{sec:results} we test each of these kernels on different examples and discuss their performance.

The kernel function is used to construct the covariance matrix $K$ of the GP. Following \cite{ambikasaran_fast_2015,rasmussen_gaussian_2006}, we add a diagonal white noise term to the kernel 
\begin{equation}
    K_{ij} = k(x_i, x_j) + \sigma_n^2 \mathbb{I},
    \label{eqn:Kij}
\end{equation}
where $K_{ij}$ are the elements of covariance matrix $K \in \mathbb{R}^{n\times n}$, $\sigma_n^2$ is the variance of the Gaussian white noise distribution, and $\mathbb{I}$ is the identity matrix.

The free parameters of the covariance matrix are known as \emph{hyperparameters} of the GP. These hyperparameters, which we denote as $\theta$ include parameters such as the length scales of each dimension, amplitude, and white noise variance, \textit{e.g.,} $\theta = \{l_1, ..., l_m, A, \sigma_n, ...\}$. In Section \ref{subsec:hyperparam_opt} we discuss different methods for tuning these free parameters.

\subsection{Training with Active Learning} \label{subsec:active_learning}

Training the GP surrogate model consists of two steps: first, an initial sampling phase in which the GP is trained on evenly sampled grid points, and second, an active learning sampling phase in which new training points are iteratively selected to optimally improve GP convergence. 

\begin{figure*}[ht!]
\begin{center}
    \includegraphics[height=11cm]{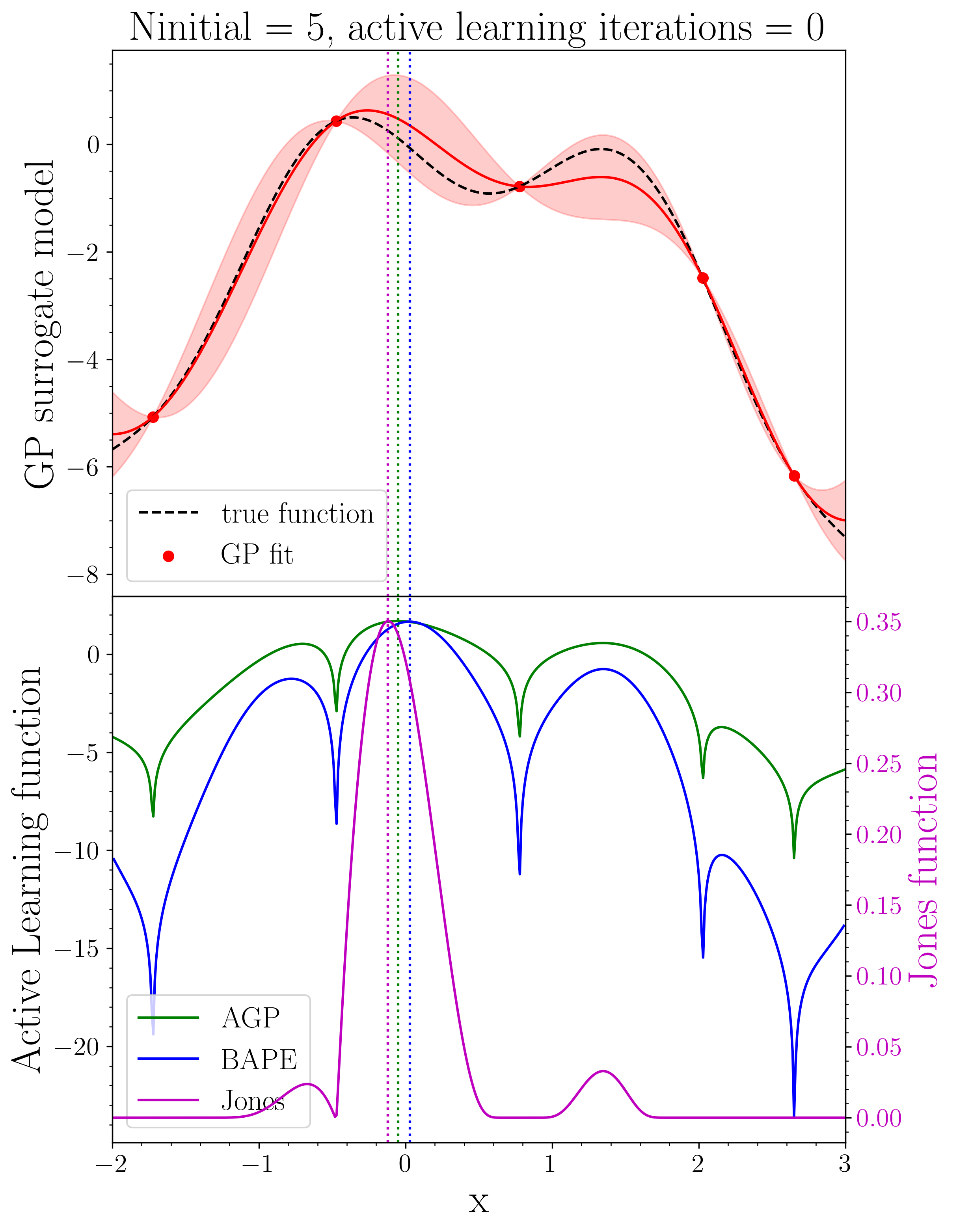} 
    \includegraphics[height=11cm]{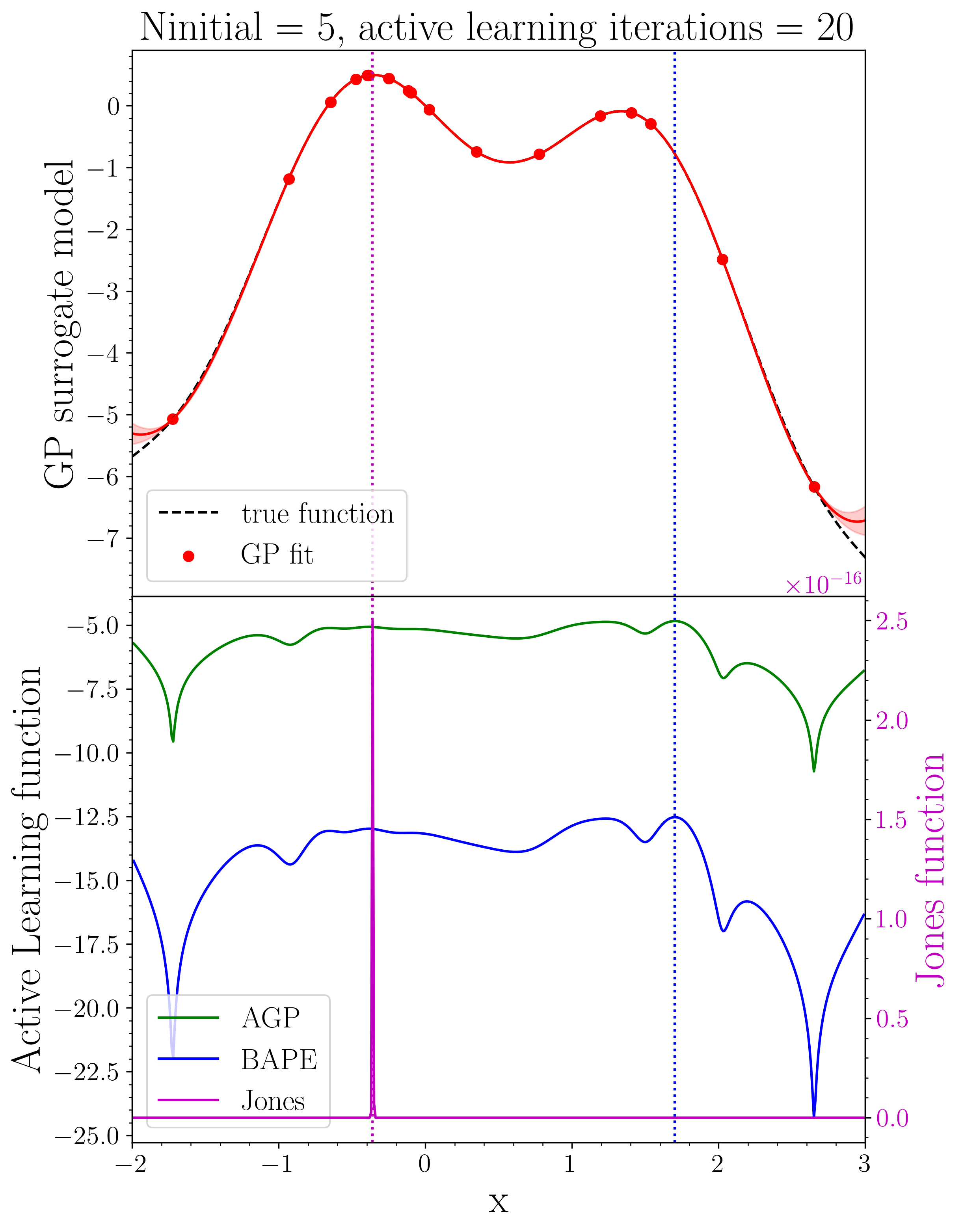} 
    \caption{Demonstration of a GP trained on a 1D probability distribution function. 
    The left panels show the GP conditioned on $N_{\rm initial}=5$ initial training points, while the right panel shows the GP conditioned on $N_{\rm initial}=5$ initial training points plus $N_{\rm active}=20$ active learning points. The top panels show the true function in black compared to the GP prediction conditioned on the training points in red, where the red line shows the GP mean prediction and the shaded red region show the 1-$\sigma$ standard deviation of the prediction. The bottom panel shows the Jones, BAPE and AGP acquisition functions Equations (\ref{eqn:jones}--\ref{eqn:agp}) as a function of input parameter $x$, with the left y-axis (in black) showing the magnitude of the BAPE and AGP functions, and the right y-axis (in purple) showing the magnitude of the Jones function. The dashed vertical lines show the maximum value of each acquisition function, indicating where the next training point will be chosen according to that algorithm.
    }
    \label{fig:active_learning}
\end{center}
\end{figure*}

New points are chosen by first evaluating the predicted output of the current iteration of the GP.
For any point in the input domain, $x^*$, the GP outputs a distribution of outcomes $y^* = \mathcal{GP}(x^*)$.  Taking the expectation values over the distribution, one can compute the predictive mean and variance at point $x^*$ as:
\begin{align}
    \mathbb{E}[y^*] &= k(x_*,\textbf{x})^T K^{-1} \mathbf{y} \label{eqn:gp_mean} \\
    \mathbb{V}{\rm ar}[y^*] &= k(x_*,x_*) - k(x_*,\textbf{x})^T K^{-1} k(x_*,\textbf{x})^T. 
    \label{eqn:gp_var}
\end{align}
To avoid confusion, $k(x_*,x_*) \in \mathbb{R}$ is a scalar with the kernel function evaluated at $x_*$, while
$k(x_*,\textbf{x}) \in \mathbb{R}^n$ is a vector between test point $x_*$ and training data $\mathbf{x}$, and $K \in \mathbb{R}^{n\times n}$ is the matrix computed with the training data $\mathbf{x}$ and white noise (Eqn. \ref{eqn:Kij}).

\alabi implements several different algorithms known as \emph{acquisition functions} to perform active learning. One of the most common acquisition functions is called the Expected Improvement \citep[EI;][]{jones_efficient_1998}, which is
\begin{equation}
    U_{\rm jones}(\textbf{x}) = (\mu_t(\textbf{x}) - f_{\rm best} - \zeta) \Phi(z) + \sigma_t(\textbf{x}) \Psi(z), \label{eqn:jones}
\end{equation}
where $z = (\mu - f_{\rm best} - \zeta) / \sigma$, $\mu_t$ and $\sigma_t^2$ are the predictive mean and variance of the GP, calculated from the kernel using Equations (\ref{eqn:gp_mean}) and (\ref{eqn:gp_var}) at the test point $\mathbf{x}$, $f_{\rm best}$ is the optimal value in the current training sample, $\Phi$ is the standard normal cumulative distribution function, and $\Psi$ is the standard normal probability distribution function. There is also a free parameter $\zeta$, which tunes the algorithm's priority for exploration vs. exploitation. A higher $\zeta (\gg 0$) value focuses on exploring a wider range of parameter space, while a lower $\zeta (\approx 0$) value focuses on converging to the local optimum. As such, a higher $\zeta$ is less susceptible to converging to a local minimum and better for finding the global maximum (important for multimodal functions), while a lower $\zeta$ is better for quicker convergence.
The Jones function is ideal for problems in which the user aims to find the local/global minimum or maximum of a computationally expensive function.

For Bayesian inference problems, \alabi\ implements acquisition functions known as Bayesian Active Learning for Posterior Estimation \citep[BAPE;][]{kandasamy_query_2017}, and Adaptive Gaussian Process \citep[AGP;][]{wang_adaptive_2018}. The BAPE and AGP acquisition functions prioritize selecting new training points where the GP has high uncertainty and high probability. The acquisition functions for the BAPE and AGP are defined as:
\begin{equation} \label{eqn:bape}
    U_{\rm bape}(\textbf{x}) = \exp[2 \mu_t(\textbf{x}) + \sigma_t^2(\textbf{x})] \cdot (\exp[\sigma_t^2(\textbf{x})] - 1)
\end{equation}
\begin{equation} \label{eqn:agp}
    U_{\rm agp}(\textbf{x}) = \mu_t(\textbf{x}) + \frac{1}{2} \log\left\vert 2\pi \exp\left[\sigma_t^2(\textbf{x}) \right] \right\vert,
\end{equation}
where $\mu$ and $\sigma^2$ are the predictive mean and variance, Equations (\ref{eqn:gp_mean}--\ref{eqn:gp_var}). 

Figure \ref{fig:active_learning} shows an example of each acquisition function applied to a GP surrogate model. Top panels show the mean and variance prediction of the GP. The bottom panels show active learning acquisition functions, where blue is the BAPE function \citep{kandasamy_query_2017}, green is the AGP function \citep{wang_adaptive_2018}, and purple is the Jones function \citep{jones_efficient_1998} plotted on a separate $y$ scale. The vertical dashed lines indicate the optimal next point to evaluate the model to improve the GP fit. While the Jones function is concentrated on the maximal function value, the BAPE and AGP functions optimize for selecting new points with high probability and high GP prediction uncertainty.

\subsection{Hyperparameter Optimization} \label{subsec:hyperparam_opt}

\alabi implements two different options for GP hyperparameter optimization: (1) marginal likelihood optimization, and (2) cross-validation.

\subsubsection{Marginal Likelihood Approach} \label{subsubsec:ml_approach}

In the marginal likelihood optimization approach, hyperparameters are chosen by solving for the set of $\theta = \{l_1, ..., l_m, A, \sigma_n, ...\}$ that maximizes the log marginal likelihood:
\begin{equation} \label{eqn:marginal}
    \log P(\textbf{y} | \textbf{x}, \theta) = 
        -\frac{1}{2} \textbf{y}^T K^{-1} \textbf{y}
        -\frac{1}{2} \log \vert K \vert
        -\frac{n}{2} \log 2\pi,
\end{equation}
where $K$ is the covariance matrix with elements given by Eqn. \ref{eqn:Kij}. The term $-\frac{1}{2} \textbf{y}^T K^{-1} \textbf{y}$ accounts for the data fit (which decreases with length scale), the term $-\frac{1}{2} \log \vert K \vert$ serves as a complexity penalty (which increases with length scale), and the term $-\frac{n}{2} \log 2\pi$ acts as a normalization constant \citep[see Section 5.4 of][]{rasmussen_gaussian_2006}. 

In principle, the complexity penalty should steer the solution away from over-fitting to the data. However, performance can be poor for high dimensions, as the length scale needed to properly condition a GP increases with the number of dimensions \citep{hvarfner2024}. This difference may not be immediately intuitive, but if we consider randomly sampled points from a unit cube, the average pairwise distance between any given points increases as the number of dimensions of the space increases. For stationary kernels, the kernel value decreases to 0 as the distance between two points increases. 

Thus, to maintain correlations in high dimensions, larger length scales are needed. To address this issue, we adopt the length scale regularization function from \cite{hvarfner2024}:
\begin{equation}
    P(\ell_i) = \log\mathcal{N} \left(\mu_{\ell} + \frac{\log(N_{\rm dim})}{2} , \sigma_{\ell} \right)
    \label{eqn:scale_reg}
\end{equation}
where $N_{\rm dim}$ is the number of dimensions of the parameter space, $\mu_\ell=2$ is the mean offset, and $\sigma_\ell=1$ is the standard deviation for the length scale distribution. \cite{hvarfner2024} show that the addition of this regularization function can improve the performance of simple stationary kernels in high dimensions without resorting to more complex kernel functions.

Optionally (for higher dimensional problems), the user can apply the \cite{hvarfner2024} length scale regularization term by setting the regularization weight $w_r > 0$. Thus, the optimization problem becomes:
\begin{equation}
    \hat{\theta}_{ml} = \underset{\theta}{\arg\max} \left[
    \underbrace{\log P(\textbf{y} | \mathbf{x}, \theta)}_{\mathrm{marginal \,\, likelihood}} + \underbrace{w_r \cdot P(\ell_i)}_{\mathrm{regularization}}
\right]
\end{equation}
using the marginal likelihood (\ref{eqn:marginal}) and the length scale regularization term (\ref{eqn:scale_reg}).
In practice, we use the \george\ \citep{ambikasaran_fast_2015} Python implementation for GP training, and optimize hyperparameters numerically using \code{scikit-optimize} \citep{head_scikit-optimizescikit-optimize_2020}. \alabi automatically makes use of the gradient function, enabling the user to use gradient-based optimization methods such as the Newton-CG algorithm \citep{gould_1999}, which can speed up the optimization by a factor $\sim100\times$ compared to non-gradient methods such as the Nelder-Mead algorithm \citep{Nelder1965}.

\subsubsection{Cross-validation Approach} \label{subsubsec:cv_approach}

Another method for optimizing GP hyperparameters is the k-fold cross-validation approach. In the k-fold cross validation approach \citep{stone_cross-validatory_1974,kohavi_1995}, the training sample is divided into $k$ evenly distributed subsamples, indexed by $j \in \{1,\dots,k \}$. To compute the best set of hyperparameters $\theta$, we select a subset $\mathbf{x}_j$ and compute the GP prediction while training on all samples not including set $j$ (which we denote as $\mathbf{x}_{/j}$). We then compute the mean squared error between the GP prediction for $\mathbf{x}_j$, and the true values $\mathbf{y}(\mathbf{x}_j)$. This calculation is repeated for all subsets $j \in \{1,\dots,k \}$, and the total cross-validation score for a given set of hyperparameters is the sum of the mean squared error over all $j$. Thus, the optimal cross-validation hyperparameters are given by:
\begin{equation}
    \hat{\theta}_{cv} = \underset{\theta}{\arg\min} \sum_{j=1}^k  \left[\mathcal{GP} (\mathbf{x}_j ; \mathbf{x}_{/j},\theta) \, - \,  \mathbf{y}(\mathbf{x}_j) \right]^2,
\end{equation}
where $\mathcal{GP} (\mathbf{x}_j ; \mathbf{x}_{/j},\theta)$ is the GP prediction for the sample $\mathbf{x}_j$ trained on dataset $\mathbf{x}_{/j}$ with hyperparameters $\theta$, and $\mathbf{y}(\mathbf{x}_j)$ is the true value. This method is often more robust at preventing over-fitting than the marginal likelihood approach (\ref{subsubsec:ml_approach}), however, it is usually more computationally expensive to perform as it requires retraining the GP $k$ times per set of hyperparameters considered.

\subsection{Scaling the Training Data} \label{subsec:training_data}


In many multi-dimensional problems, the ranges of interest for each parameter may have very different length scales, which can sometimes be orders of magnitude different. To better condition the data set for the GP fitting, \alabi implements different scaling options. By default \alabi uses a minmax scaler to scale training inputs to the range [0,1] when training the GP, and implements transform functions so that the user can easily call the GP model in real coordinates.

Another documented issue for training GPs in high dimensions is the boundary sampling issue, in which the acquisition function optimization (\ref{subsec:active_learning}) tends to over sample the edges of the prior boundaries \citep{Swersky2017}. To account for cases in which boundary clustering becomes an issue with active learning, \alabi\ implements the option for applying the beta warping function from \cite{Swersky2017}:
\begin{equation}
    w_i(x_i) = \int_0^{x_i} \frac{u^{\alpha-1} (1 - u)^{\beta-1}}{\mathcal{B}(\alpha,\beta)} du 
    \label{eqn:beta_scaling},
\end{equation}
where $\mathcal{B}$ is the beta distribution with scale parameters $\alpha$ and $\beta$.

\subsection{Markov Chain Monte Carlo} \label{subsec:mcmc}

\alabi\ is modularized to interface with different MCMC packages for sampling posterior distributions, given any black-box likelihood function. 
Currently \alabi\ provides integrated support for the \emcee\ affine-invariant sampling package \citep{foreman-mackey_emcee_2013}, as well as the nested sampling \citep{Skilling2004} packages \dynesty\ \citep{speagle_dynesty_2020} and \multinest\ \citep{feroz_multinest_2009,Buchner2014} and \ultranest\ \citep{Buchner2021}. For a detailed description of each algorithm, see their respective papers. \alabi\ provides a unified interface for conveniently testing different samplers on their problem. 

\subsection{Performance Metrics} \label{subsec:metrics}

The performance of \alabi\ depends on how well the GP surrogate model converges to the true model after training the model with active learning. To quantify how well the surrogate model is performing, we compute the Kullback-Leibler Divergence and test error metrics to assess the convergence between the surrogate and true model as a function of active learning iteration.

\subsubsection{Kullback-Leibler Divergence} \label{subsubsec:kl_div}

The dissimilarity between two probability distributions $P$ and $Q$ can be quantified by the Kullback-Leibler (KL) divergence metric:
\begin{equation}
    D_{\rm KL}(P \Vert Q) = \int_{\mathcal{D}} p(x) \log \left\vert \frac{p(x)}{q(x)} \right\vert \, dx
    \geq 0,
    \label{eqn:kl_div}
\end{equation}
where $P(x)$ and $Q(x)$ represent two probability density functions, and the integral is taken over the prior domain $\mathcal{D}$. 
The KL divergence measures the average difference in the logarithm of the ratio of probabilities between the two distributions. It indicates how much information is lost when using $Q$ to approximate $P$. If $D_{\rm KL}(P \Vert Q) = 0$, it implies that the two distributions are identical, whereas a larger KL divergence value indicates greater dissimilarity. 

When the distributions $P(x) \sim \mathcal{N}(\mu_1, K_1)$ and $Q(x) \sim \mathcal{N}(\mu_2, K_2)$ are both Gaussian distributions, KL divergence between two $P(x)$ and $Q(x)$ can be written as:
\begin{equation}
\begin{split}
    D_{\rm KL}(P \Vert Q) = 
     \frac{1}{2} \left[ \log \frac{\vert K_2 \vert}{\vert K_1 \vert} 
        + \mu^T K_2^{-1} \mu 
        + \mathrm{tr} \left(K_2^{-1} K_1\right) - k
    \right],
    \label{eqn:kl_div_gaussians}
\end{split}
\end{equation}
where $K_1$ is the covariance matrix of $P$, $K_2$ is the covariance matrix of $Q$, and $\mu = \mu_1 - \mu_2$ is the difference in the mean of $P$ and $Q$. We apply this KL divergence metric in Section \ref{sec:results} to benchmark how well sampling with \alabi\ compares to the ground truth.

For the general case, when the distributions $P$ and/or $Q$ are not Gaussian, there are a variety of different methods for estimating the KL divergence, which typically involve computing $D_{\rm KL}(P \Vert Q)$ using density estimates of P and Q (see \citealt{perez-cruz_kullback-leibler_2008} and the citations therein). Following this approach, we numerically estimate the KL divergence by first sampling the distributions using MCMC (\ref{subsec:mcmc}) to get samples ${x_i} \sim p(x)$. We then use kernel density estimation \citep{Rosenblatt1956,Parzen1962} to get the estimated density functions $\tilde{p}(x)$ and $\tilde{q}(x)$, and estimate the integral in Equation (\ref{eqn:kl_div}) as the sum over all samples $x_i$:
\begin{align}
    \hat{D}_{\rm KL}(P \Vert Q) &= \sum_{x_i \in \mathcal{D}} \tilde{p}(x_i)  \log \left\vert \frac{\tilde{p}(x_i)}{\tilde{q}(x_i)} \right\vert  \\
        &= \mathbb{E}_{x_i}[\log \tilde{p}(x_i) - \log \tilde{q}(x_i)] \nonumber,
\end{align}
which is equivalent to taking the mean over the difference in log distributions. \\

\subsubsection{Test Sample Error} \label{subsubsec:test_error}

In most Bayesian inference problems that one would want to use \alabi for, it would be impractical to compute the KL divergence between the true distribution and the surrogate model, as it would either require having an analytic function for the likelihood, or require sampling the true function. As a more practical measure of convergence, \alabi also includes the option of computing the \emph{test error}. These test points are computed from the true model, but are held out from the training sample so that the accuracy of the surrogate model can be quantified while not being biased by overfitting the GP. For example, given a set of $N_{\rm test}$ test samples, we can compute the mean squared error as:
\begin{equation}
    \hat{\Delta}_{\rm mse} = \frac{1}{N_{\rm test}} \sum_{i=1}^{N_{\rm test}} \left( \frac{y_{\rm gp}(x_i) - y_{\rm true}(x_i)}{y_{\rm true}(x_i)} \right)^2,
    \label{eqn:mse}
\end{equation}
or we can compute the average percent error as:
\begin{equation}
    \hat{\Delta}_{\rm prc} = \frac{1}{N_{\rm test}} \sum_{i=1}^{N_{\rm test}} \left| \frac{y_{\rm gp}(x_i) - y_{\rm true}(x_i)}{y_{\rm true}(x_i)} \right| \times 100,
    \label{eqn:percent_error}
\end{equation}
where $y_{\rm true}(x_i)$ is the prediction from the true model and $y_{\rm gp}(x_i)$ is the prediction from the surrogate model at each point $x_i$ for $i=1,\dots, N_{\rm test}$ in the test sample. \\

\section{Implementation} \label{sec:implement}

\subsection{Basic Example} \label{subsec:basic}

The \alabi\ package builds off of previous work known as \approxposterior\ \citep{fleming_approxposterior_2018}, with improved modularity, interface with additional sampling packages, as well as user-friendly features for saving and benchmarking results. In this section we provide a minimum use case example to familiarize users with the \alabi\ workflow.

First the user defines a function to train the GP surrogate model. For the case of a Bayesian inference problem this function would represent the posterior, $\ln\mathrm{Posterior} \propto \ln\mathrm{Likelihood} + \ln\mathrm{Prior}$. As a simple example, we define a 1-dimensional multimodal function to train (Figure \ref{fig:active_learning}):
\begin{lstlisting}[gobble=4]
    import numpy as np
    
    def lnp(x):
        d1 = 0.6 * np.exp(-(x + 2)**2/2)
        d2 = 0.4 * np.exp(-(x - 2)**2/2)
        return np.log((d1 + d2)/np.sqrt(2*np.pi))

\end{lstlisting}

The next step is to train the GP surrogate model using \alabi. In this example, we start with an initial training sample of \code{ntrain=10} points, evenly sampled from the prior space using the Sobol sampler (\code{sampler="sobol"}).

\begin{lstlisting}[gobble=4]
    import alabi
    from alabi import SurrogateModel

    kernel = "ExpSquaredKernel"
    savedir = f"results/{kernel}"
    
    sm = SurrogateModel(lnlike_fn=lnp, 
                        bounds=[(-5,5)], 
                        savedir=savedir)

    # Randomly sample 10 initial training points 
    sm.init_samples(ntrain=10, sampler="sobol")
\end{lstlisting}

We then initialize the GP model by fitting it to the initial training sample points. At this stage, we can specify the kernel function (e.g., the exponential squared kernel, Eqn. \ref{eqn:exp_squared_kernel}) as well as the additional hyperparameters (in this case an amplitude and mean parameter). \alabi\ also has the option to set a white noise parameter, where the variable \code{white\_noise} sets the logarithm of the noise scale. The white noise option adds uncorrelated noise along the diagonal of the covariance matrix (Eqn. \ref{eqn:Kij}) that improves the numerical stability of computing the GP. 

\begin{lstlisting}[gobble=4]
    # Fit the GP to the initial training sample
    sm.init_gp(kernel=kernel, 
               fit_amp=True, 
               fit_mean=True, 
               white_noise=-12)
\end{lstlisting}

After the GP surrogate model is trained on an initial sample of training points, the next step is to iteratively sample new training points using active learning. 
At this stage, the user can specify the number of iterations, \code{niter} and active learning algorithm, \code{algorithm} (with options including \code{"bape"}, \code{"agp"}, or \code{"jones"}). The \code{save\_progress} option (set to \code{True} or \code{False}) allows the user to automatically cache results to a pickle file each time the hyperparameters are optimized, which is useful if your computation may be interrupted or restarted.  
\begin{lstlisting}[gobble=4]
    # Run the BAPE active learning algorithm for 20 iterations
    sm.active_train(niter=20, algorithm="bape", gp_opt_freq=5, save_progress=True)
    sm.plot(plots=["gp_all"])
\end{lstlisting}
As the GP trains on more data points, \alabi\ has the option to improve the fit by re-optimizing the hyperparameters after a specified number of iterations set by \code{gp\_opt\_freq}. In this example we re-optimize the hyperparameters every 5 active learning iterations.

After training the GP, the MCMC sampler can be run to sample the surrogate model, e.g., using the \emcee and/or \dynesty package:
\begin{lstlisting}[gobble=4]
    # Run MCMC sampler with emcee
    sm.run_emcee(nwalkers=10, nsteps=int(5e4))
    sm.plot(plots=["emcee_all"])

    # Run MCMC sampler with dynesty
    sm.run_dynesty()
    sm.plot(plots=["dynesty_all"])
\end{lstlisting}
At this stage, the user can define the prior function that they want to use to sample their posterior. This modularity allows for the user to test different prior assumptions without retraining the surrogate model on the likelihood function. If the prior function is left unspecified, \alabi will default to using a uniform prior.

\begin{figure}[ht!]
\begin{center}
    \includegraphics[width=\linewidth]{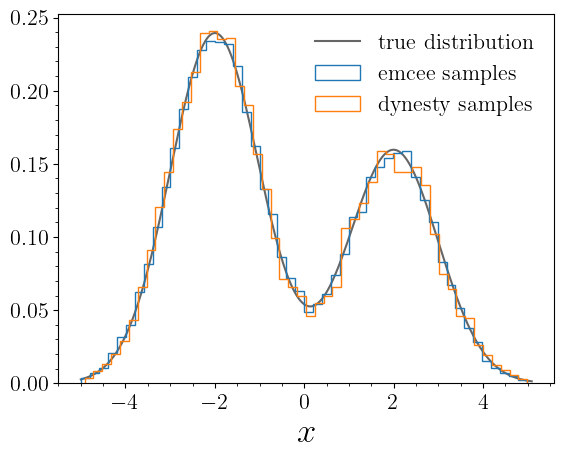} 
    \caption{Probability density of the true distribution (black) compared to the MCMC samples drawn from the GP surrogate model after 10 initial training samples + 20 active learning training samples with \code{emcee} (blue) and \code{dynesty} (orange).}
    \label{fig:mcmc_demo}
\end{center}
\end{figure}

Figure \ref{fig:mcmc_demo} shows the probability density of the true distribution compared to the sampling with the GP surrogate model with \emcee\ and \dynesty.
Results are automatically saved to a pickle file that can be loaded and rerun:
\begin{lstlisting}[gobble=4]
    sm = alabi.cache_utils.load_model_cache(savedir)
\end{lstlisting}

\alabi\ also saves a summary of each training run in a human-readable text file that summarizes the configuration parameters used.  \\


\begin{figure*}[ht]
    \centering
    \includegraphics[width=0.75\linewidth]{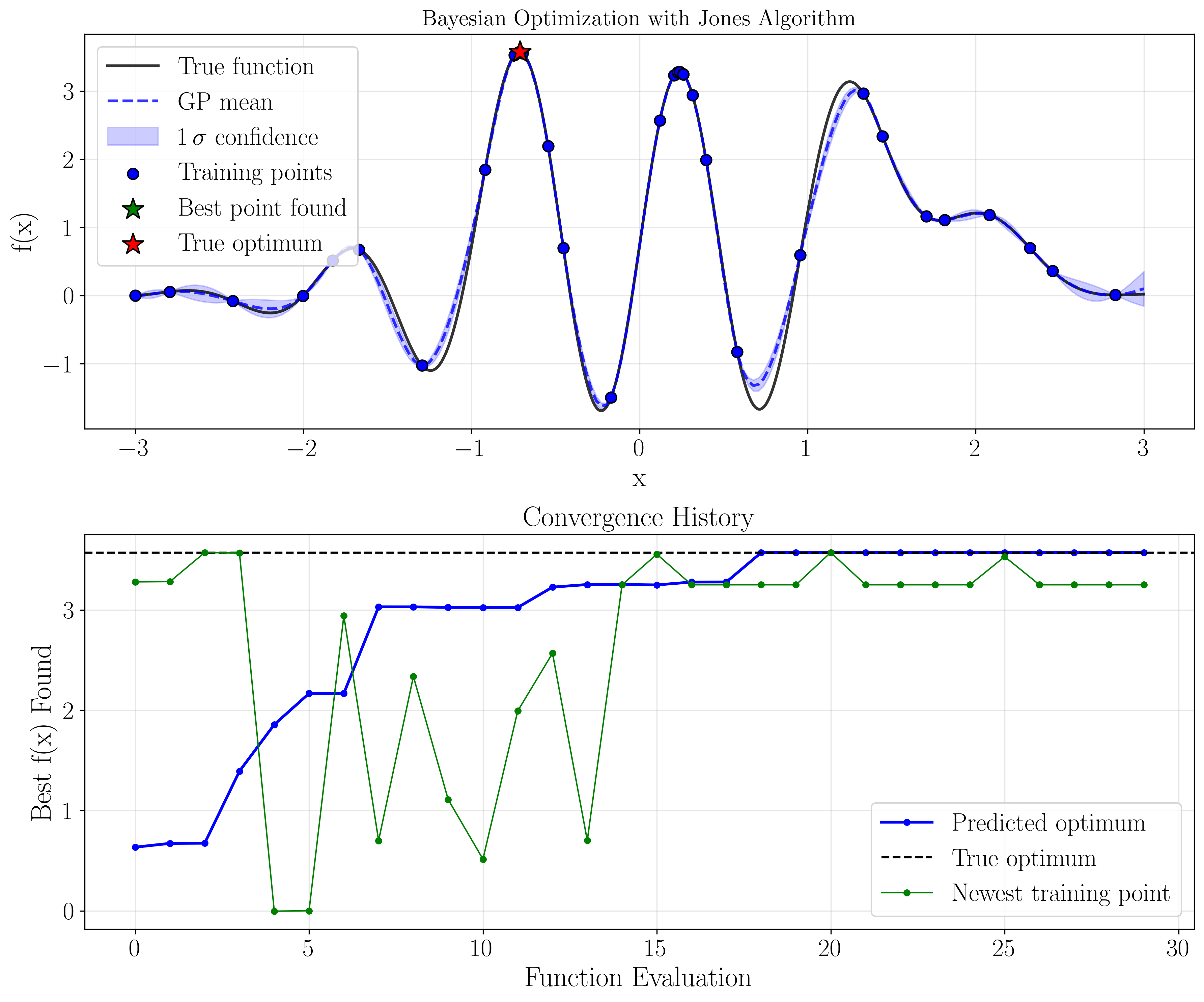}
    \caption{An example using \alabi\ for Bayesian optimization on a 1-dimensional multimodal function. The top panel shows the GP prediction and training points compared to the true function.  The bottom panel shows the convergence history of the GP as a function of iteration. \\}
    \label{fig:bayes_opt}
\end{figure*}

\subsection{Advanced Usage} \label{subsec:advanced}

For more advanced usage, there are a number of different settings and hyperparameters of the training process that can be adjusted to improve performance according to the user's problem, for example, the choices of: kernel function, the active learning algorithm, the number of initial or active learning training points, or the MCMC sampler. 
\alabi\ also includes functionality for running the GP training and MCMC sampling in parallel to make use of multi-core hardware. To see examples and discussion of advanced use cases (such as parallelization, defining custom priors, or creating diagnostic plots), see the most recent documentation\footnote{\href{https://jbirky.github.io/alabi/build/index.html}{https://jbirky.github.io/alabi/build/index.html}}. \\


\section{Results and Performance} \label{sec:results}

In this section, we provide some illustrative examples of how different \alabi\ configuration settings perform on a variety of test cases, including distributions with degeneracies, multimodalities, and high dimensionality. 
It is not always easy to determine which settings are most appropriate for a given problem. However, in the following section, we present several illustrative examples for a variety of test cases to provide an intuition for how to select different settings in practice.
We furthermore demonstrate the computational efficiency of these methods in terms of computation time and number of forward model evaluations.
We also compare the performance of the \emcee, \dynesty, \ultranest, and \multinest\ MCMC sampling algorithms for each of the test cases.

In order to assess \alabi's performance on different test cases, we show how \alabi\ performs against several benchmarks where we know the ground truth. In most real-life use cases, the user does not know the ground truth for their posterior function. This section is designed to provide a heuristic intuition for how \alabi performs in different scenarios so that the user can choose the training settings (e.g., number of training iterations, kernel function, MCMC sampler) that are appropriate for their problem. 


\subsection{Bayesian Optimization} \label{subsec:bayes_opt}

Bayesian optimization is a common application for this framework. Figure \ref{fig:bayes_opt} shows how \alabi can be used to maximize a multimodal peaks function.
As seen in the top plot, points sampled by the Jones active learning algorithm are clustered around maximum function values to improve the precision of the optimum value. In this example, the GP optimum converges to the true optimum within 20 iterations.

\clearpage
\subsection{Benchmark Problems} \label{subsec:benchmarks}


\begin{figure}[!ht]
    \centering
    \includegraphics[width=0.5\textwidth]{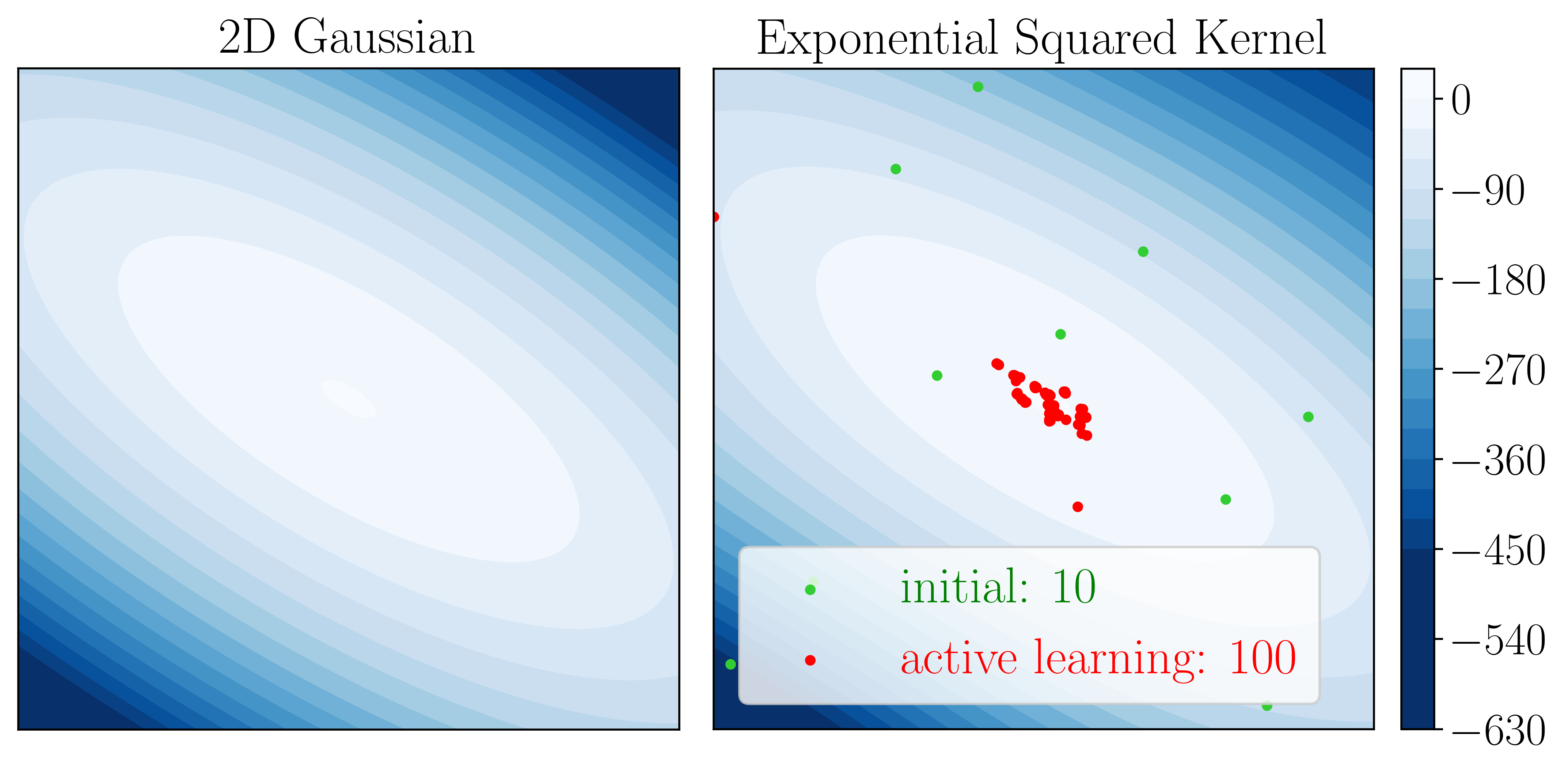}
    \caption{Two-dimensional Gaussian function (Example \ref{subsubsec:gaussian_2d}) with the true function (left panel) compared to the GP surrogate model (right panel). The green points highlight the quasi-uniform sampled initial training points, and the red points highlight training points sampled iteratively using active learning.} 
    \label{fig:example_start}
    \includegraphics[width=0.5\textwidth]{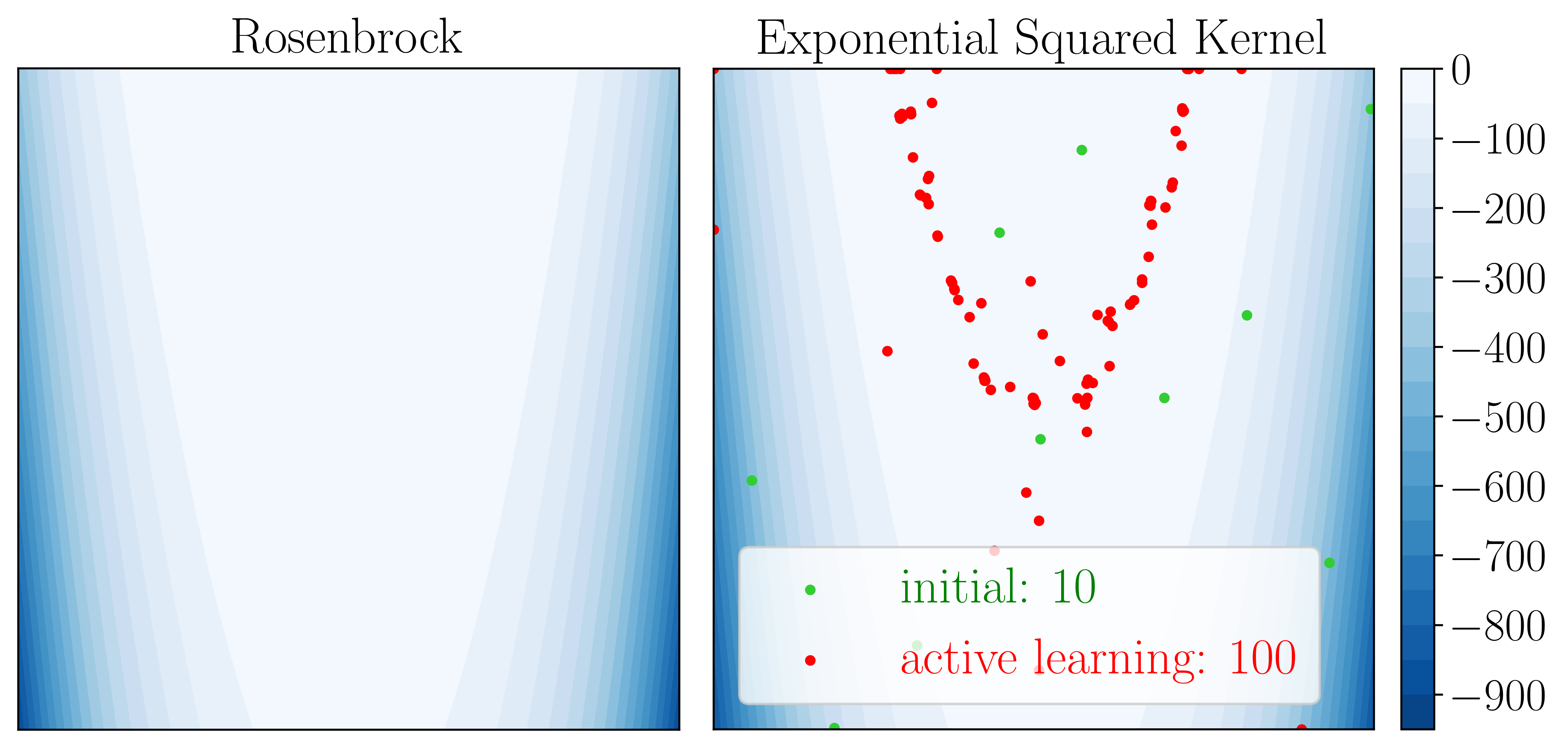}
    \caption{Rosenbrock function (Example \ref{subsubsec:rosenbrock}).}
    \label{fig:rosenbrock}
    \includegraphics[width=0.5\textwidth]{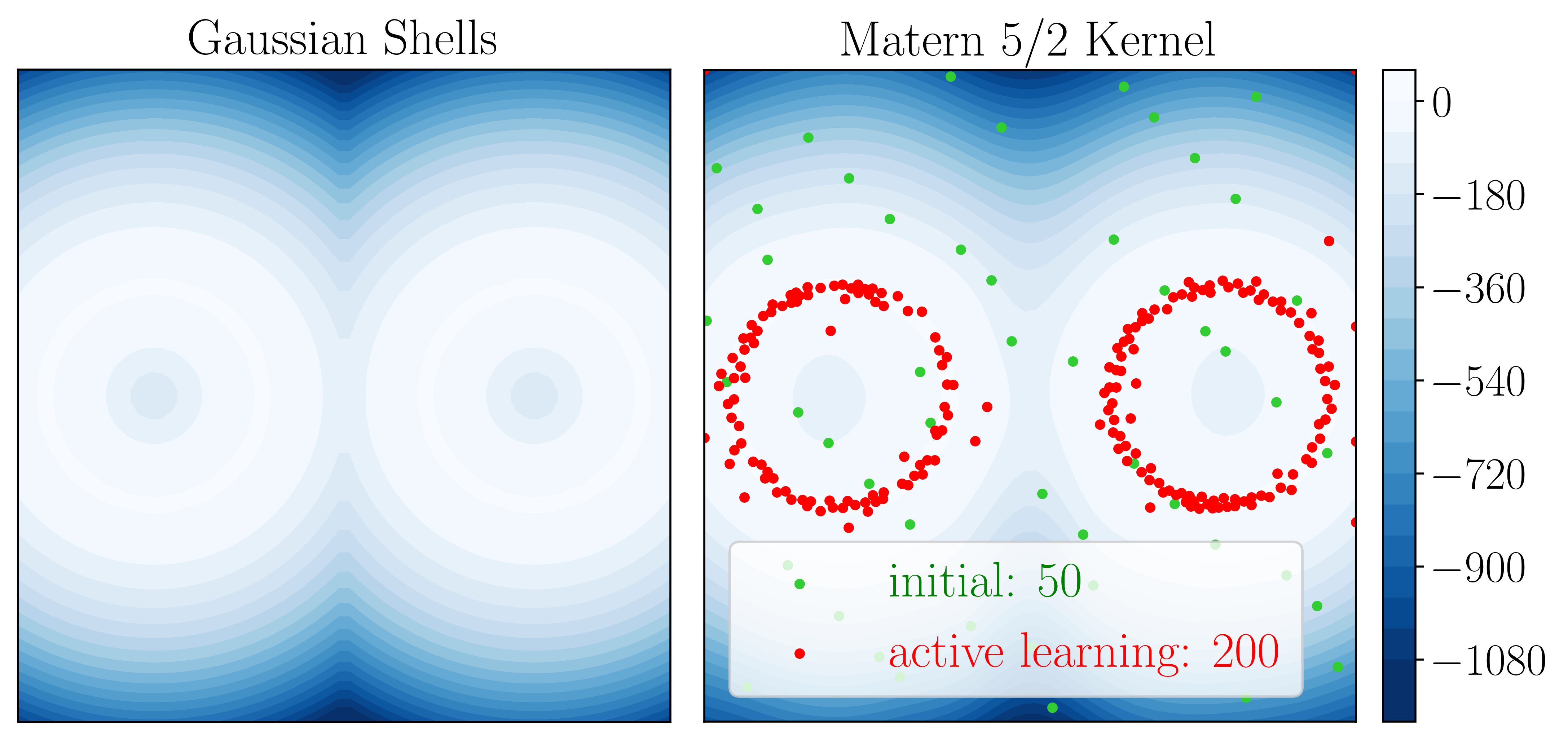}
    \caption{Gaussian shells function (Example \ref{subsubsec:gaussian_shells}).}
    \label{fig:gaussian_shells}
    \includegraphics[width=0.5\textwidth]{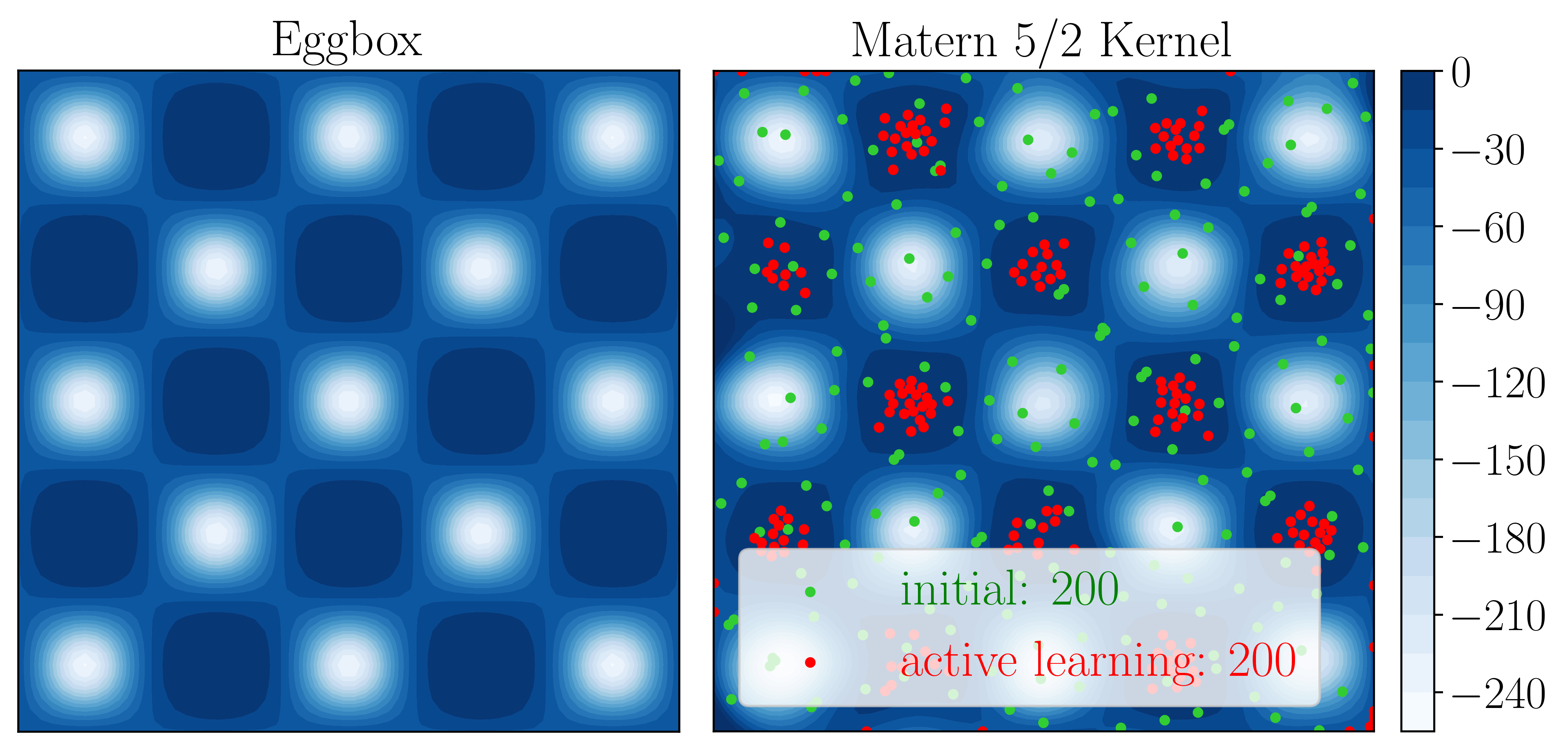}
    \caption{Eggbox function (Example \ref{subsubsec:eggbox}).}
    \label{fig:example_end}
\end{figure}


\subsubsection{Example 1: 2-dimensional Gaussian} \label{subsubsec:gaussian_2d}

As a first example, we test \alabi on a simple unimodal Gaussian distribution, with mean and covariance parameters
\begin{equation}
    f_1(\textbf{x}) = \mathcal{N}\left(
    \begin{bmatrix} 0 \\ 0 \end{bmatrix}, 
    \begin{bmatrix} 0.2 & -0.1 \\ -0.1 & 0.1 \end{bmatrix}
    \right).
\end{equation}
Figure \ref{fig:example_start} shows the true function compared to the GP surrogate model trained with 10 initial training samples, and 100 active learning samples. For a simple unimodal distribution, not many training points are needed to replicate the true function, which we show quantitatively in Section \ref{subsec:2d_performance}.

\subsubsection{Example 2: 2-dimensional Rosenbrock function} \label{subsubsec:rosenbrock}

As a second example, we test the GP performance on a 2D function with a curved degeneracy feature given in \cite{wang_adaptive_2018}:
\begin{equation} 
    f_2(\textbf{x}) = \exp\left[-\frac{1}{100} (x_1 - 1)^2
        - (x_1^2 - x_2)^2 \right],
    \label{eqn:rosenbrock}
\end{equation}
with the input defined over the prior domain $x_i \in [-5,5]$. Figure \ref{fig:rosenbrock} shows the true function compared to the GP surrogate model trained with 10 initial training samples, and 100 active learning samples.

\subsubsection{Example 3: 2-dimensional Gaussian shells} \label{subsubsec:gaussian_shells}

As a third example, we test how the GP performs with highly correlated features using the Gaussian shells example from \cite{speagle_dynesty_2020}:
\begin{equation}
    f_3(\mathbf{x}) = \mathcal{C}(\mathbf{x} | c_1, r_1, w_1) + \mathcal{C}(\mathbf{x} | c_2, r_2, w_2),
    \label{eqn:gaussian_shells}
\end{equation}
with the input defined over the domain $x_i \in [-6,6]$ and each circle is given by the distribution:
\begin{equation}
    \mathcal{C}(\mathbf{x} | c, r, w) 
        = \frac{1}{\sqrt{2\pi w^2}} 
        \exp \left[-\frac{1}{2} \frac{(|\mathbf{x} - c| - r)^2}{w^2} \right].
\end{equation}
Figure \ref{fig:gaussian_shells} shows the true function compared to the GP surrogate model trained with 50 initial training samples, and 200 active learning samples. Compared to the unimodal Gaussian and Rosenbrock functions, the shape of the Gaussian shells function is more complex and requires a greater number of training points to converge to the true model.

\subsubsection{Example 4: 2-dimensional multimodal} \label{subsubsec:eggbox}

As a fourth example, we test the GP performance on a highly multimodal 2D distribution known as the eggbox function \citep{feroz_importance_2019}:
\begin{equation}
    f_4(\mathbf{x}) = \exp \left\{ \left[2 + \cos\left(\frac{x_1}{2}\right) \cos\left(\frac{x_2}{2} \right) \right]^5 \right\}
    \label{eqn:eggbox}
\end{equation}
with the input defined over the domain $x_i \in [0,1]$.
Figure \ref{fig:example_end} shows the true function compared to the GP surrogate model trained with 100 initial training samples, and 250 active learning samples. Out of all of the 2D examples we show, the eggbox function is the most difficult to model and takes the most number of training points to converge.


\subsection{MCMC Results on the True Posteriors} \label{subsec:mcmc_true}

\alabi\ provides a flexible, and easy to use interface to switch between different popular MCMC samplers.
We first see how each MCMC sampler performs when sampling the true functions. 
For this test, we use the configuration settings from Table \ref{tab:mcmc_config} as the key word options for each sampler, in the following example:

\begin{lstlisting}[gobble=4]
    from alabi import SurrogateModel
    from alabi import utility as ut
    from functools import partial
    from scipy.stats import multivariate_normal

    gaussian_2d_fn = partial(multivariate_normal.logpdf, mean=np.zeros(2), cov=np.eye(2)*0.1)
    bounds = [(-3, 3) for _ in range(2)]

    # Set up uniform prior function (emcee format)
    prior_fn = partial(ut.lnprior_uniform, bounds=bounds)

    # Set up uniform prior transform (dynesty/ultranest/pymultinest format)
    prior_transform = partial(ut.prior_transform_uniform, bounds=bounds)

    # Use alabi as a wrapper for running different MCMC samplers 
    sm = SurrogateModel(gaussian_2d_fn, bounds=bounds)

    sm.run_emcee(like_fn=sm.true_log_likelihood, prior_fn=prior_fn, **emcee_kwargs)

    sm.run_dynesty(like_fn=sm.true_log_likelihood, prior_transform=prior_transform, sampler_kwargs=dynesty_sampler_kwargs, run_kwargs=dynesty_run_kwargs)
    
    sm.run_ultranest(like_fn=sm.true_log_likelihood, prior_transform=prior_transform,
    sampler_kwargs=ultranest_sampler_kwargs, run_kwargs=ultranest_run_kwargs,)
    
    sm.run_pymultinest(like_fn=sm.true_log_likelihood, prior_transform=prior_transform, sampler_kwargs=pymultinest_kwargs)
\end{lstlisting}

Table \ref{tab:mcmc_summary} summarizes how each MCMC sampler performs in sampling different functions in terms of computational runtime, evidence error, number of effective posterior samples, and sampling efficiency. 
Each of the nested samplers was run with 1000 live points and run until an evidence error of dlogz $< 0.5$ was achieved. 

Each of the samplers perform well at sampling unimodal distributions (2D Gaussian and rosenbrock), however only the nested samplers (\dynesty, \ultranest, and \multinest) converged for the multimodal examples (Gaussian shells and eggbox), while \emcee\ failed to converge, even running for longer chains of $1\times10^5$ steps. 
In terms of total computation time, \multinest generally performed the quickest of all samplers, but performed similarly to the other samplers in terms of the efficiency of computing the number of effective samples per second.

\begin{table}[h!]
    \centering
    \caption{MCMC sampler configuration parameters}
    \label{tab:mcmc_config}
    \begin{tabular}{|l|l|l|}
    \hline
    \textbf{Description} & \textbf{Parameter} & \textbf{Value} \\
    \hline
    \hline
    \multicolumn{3}{|c|}{\code{emcee}} \\
    \hline
    Number of walkers & \code{nwalkers} & 10 \\
    Steps per walker & \code{nsteps} & $1\times10^4$ \\
    Burn-in steps & \code{burn} & $1\times10^3$ \\
    \hline
    \multicolumn{3}{|c|}{\code{dynesty}} \\
    \hline
    Number of live points & \code{nlive} & 1,000 \\
    Bounding method & \code{bound} & multi \\
    Sampling method & \code{sample} & auto \\
    Posterior weight fraction & \code{pfrac} & 1.0 \\
    Maximum iterations & \code{maxiter} & $2\times10^4$ \\
    Initial evidence tolerance & \code{dlogz\_init} & 0.5 \\
    \hline
    \multicolumn{3}{|c|}{\code{UltraNest}} \\
    \hline
    Number of live points & \code{min\_num\_live\_points} & 1,000 \\
    Target effective samples & \code{min\_ess} & 2,000 \\
    Evidence tolerance & \code{dlogz} & 0.5 \\
    KL divergence tolerance & \code{dKL} & 0.5 \\
    \hline
    \multicolumn{3}{|c|}{\code{PyMultiNest}} \\
    \hline
    Number of live points & \code{n\_live\_points} & 1,000 \\
    Evidence tolerance & \code{evidence\_tolerance} & 0.5 \\
    Sampling efficiency & \code{sampling\_efficiency} & 0.8 \\
    Multimodal detection & \code{multimodal} & True \\
    Maximum modes & \code{max\_modes} & 20 \\
    \hline
    \end{tabular}
\end{table}

\begin{table*}
    \caption{MCMC Sampler Performance Summary}
    \label{tab:mcmc_summary}
    \begin{tabular}{|l|c|c|c|c|}
    \hline
    \textbf{Sampler} & \textbf{gaussian\_2d} & \textbf{rosenbrock} & \textbf{gaussian\_shells} & \textbf{eggbox} \\
    \hline
    \hline
    \multicolumn{5}{|c|}{\textbf{Runtime (seconds)}} \\
    \hline
    emcee & 13.7 & 6.0 & 5.6 & 4.2 \\
    dynesty & 25.3 & 12.6 & 61.0 & 13.5 \\
    ultranest & 15.3 & 13.8 & 18.5 & 15.1 \\
    pymultinest & 3.9 & 4.2 & 3.7 & 3.3 \\
    \hline
    \multicolumn{5}{|c|}{\textbf{Log Evidence (logZ ± error)}} \\
    \hline
    emcee & — & — & — & — \\
    dynesty & -3.65 ± 0.04 & -2.55 ± 0.03 & -1.65 ± 0.03 & -4.19 ± 0.03 \\
    ultranest & -3.62 ± 0.12 & -2.50 ± 0.07 & -1.75 ± 0.08 & -4.18 ± 0.08 \\
    pymultinest & -3.58 ± 0.01 & -2.53 ± 0.01 & -1.71 ± 0.01 & -3.81 ± 0.03 \\
    \hline
    \multicolumn{5}{|c|}{\textbf{Effective Samples}} \\
    \hline
    emcee & 5,290 & 4,090 & 550 & 590 \\
    dynesty & 15,860 & 14,820 & 15,270 & 14,631 \\
    ultranest & 9,686 & 8,100 & 7,465 & 8,777 \\
    pymultinest & 3,474 & 2,790 & 2,820 & 3,153 \\
    \hline
    \multicolumn{5}{|c|}{\textbf{Sampling Efficiency (samples/sec)}} \\
    \hline
    emcee & 386 & 684 & 98.9 & 142 \\
    dynesty & 627 & 1.2k & 251 & 1.1k \\
    ultranest & 634 & 588 & 403 & 583 \\
    pymultinest & 891 & 667 & 755 & 960 \\
    \hline
    \end{tabular}
\end{table*}

\subsection{GP performance with degeneracies and multimodalities} \label{subsec:2d_performance}

\begin{figure}[ht!]
    \centering
    \includegraphics[width=0.48\textwidth]{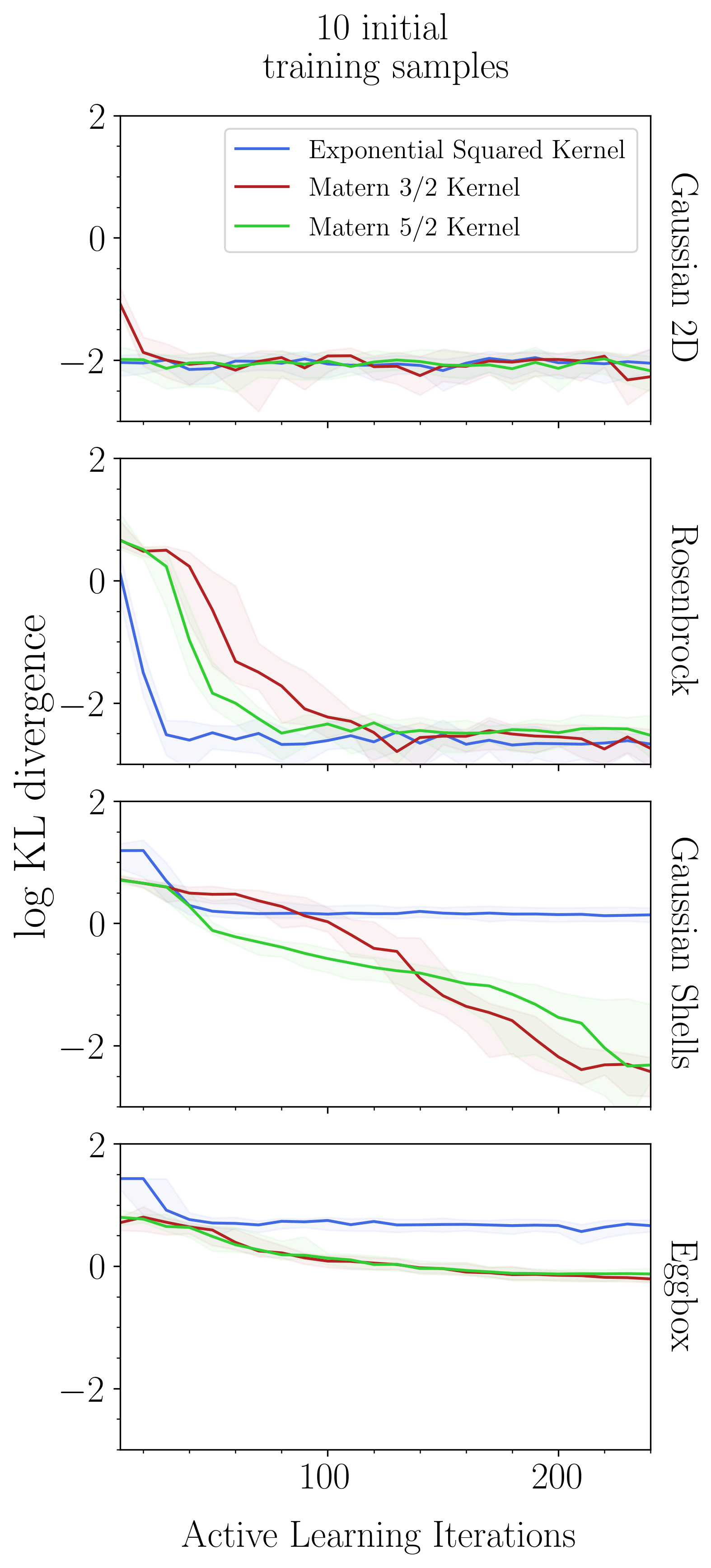}
    \caption{Performance of different GP kernel functions (exponential squared, Matern-3/2, and Matern-5/2) for each of the four different 2-dimensional examples (the Gaussian 2D, rosenbrock, gaussian shells, and eggbox functions), using \dynesty to sample each function. The y-axis shows the KL divergence metric computed between the GP surrogate model and the true model, where values closer to zero mean better convergence.}
    \label{fig:kl_vs_iteration}
\end{figure}

Next we use the true posteriors (computed in the previous section \ref{subsec:mcmc_true}) as a baseline for comparing the performance of GP surrogate models.
We explore how well the GP converges for the different 2-dimensional examples (Examples \ref{subsubsec:gaussian_2d}--\ref{subsubsec:eggbox}), by computing the KL divergence (\ref{eqn:kl_div}) between the true distribution and GP surrogate distribution.
Figure \ref{fig:kl_vs_iteration} shows how well the surrogate performs on the benchmark examples in Section \ref{subsec:benchmarks}. Each GP surrogate model was trained using 10 initial training samples, and run for 250 active learning iterations. The colors denote three different kernels: Exponential Squared (blue; Eq. \ref{eqn:exp_squared_kernel}), Matern 3/2 (red; Eq. \ref{eqn:matern32_kernel}), and Matern 5/2 (green; Eq. \ref{eqn:matern52_kernel}). Each line shows the median KL divergence value over 30 random trials with the same training configuration, and the upper and lower shaded regions show the 25th and 75th percentiles.

For the 2D Gaussian example, all kernels perform similarly well. The GP converges quickly with just the 10 initial samples for the exponential squared and Matern-5/2 kernels, and within $\sim10$ active learning iterations for the Matern-3/2 kernel. For the Rosenbrock example, the exponential squared kernel performs the best (converging within $\sim20$ active learning samples), while the Matern-3/2 kernel performs the worst (taking $\sim120$ active learning iterations to converge). 

For the last two examples, we find that the kernel performance is the opposite: the Matern-3/2 kernel performs the best, while the exponential squared kernel plateaus and does not achieve the same performance even when adding more active learning samples. This difference is due to the profile of the kernel function. Referring back to Figure \ref{fig:kernel_comparison}, the Matern-3/2 kernel has a steeper peak at small separations, whereas the exponential squared kernel is flatter at small separations, and the Matern-5/2 kernel falls somewhere between the two. This steeper peak of the Matern-3/2 kernel performs better at reproducing functions with tightly correlated modes or features (which we see in the performance of the Gaussian shells and eggbox example), while the flatter exponential squared kernel performs better on functions with broader features (which we see in the 2D gaussian and Rosenbrock examples).

\subsection{GP performance scaling with number of dimensions} \label{subsubsec:nd_gaussian}


We next test how well a GP surrogate model can replicate functions of higher dimensions by training the surrogate on an N-dimensional Gaussian distribution:
\begin{equation}
    f_5(\mathbf{x}) = \frac{1}{\sqrt{(2\pi)^{N_{\rm dim}} \det\mathbf{\Sigma}}}
    \exp \left[-\frac{1}{2} (\mathbf{x} - \mu)^T \mathbf{\Sigma}^{-1} (\mathbf{x} - \mu) \right],
    \label{eqn:nd_gaussian}
\end{equation}
where $N_{\rm dim}$ is the number of dimensions, $\mu$ is the mean vector, and $\mathbf{\Sigma}$ is the covariance matrix.  For the true parameters of the N-dimensional Gaussian, we choose a mean vector $\mu$ centered at zero in each dimension, and generate a random covariance matrix with elements $\Sigma_{ij}$ such that $\mathbf{\Sigma}$ is positive definite. 

\begin{figure}
    \includegraphics[width=\linewidth]{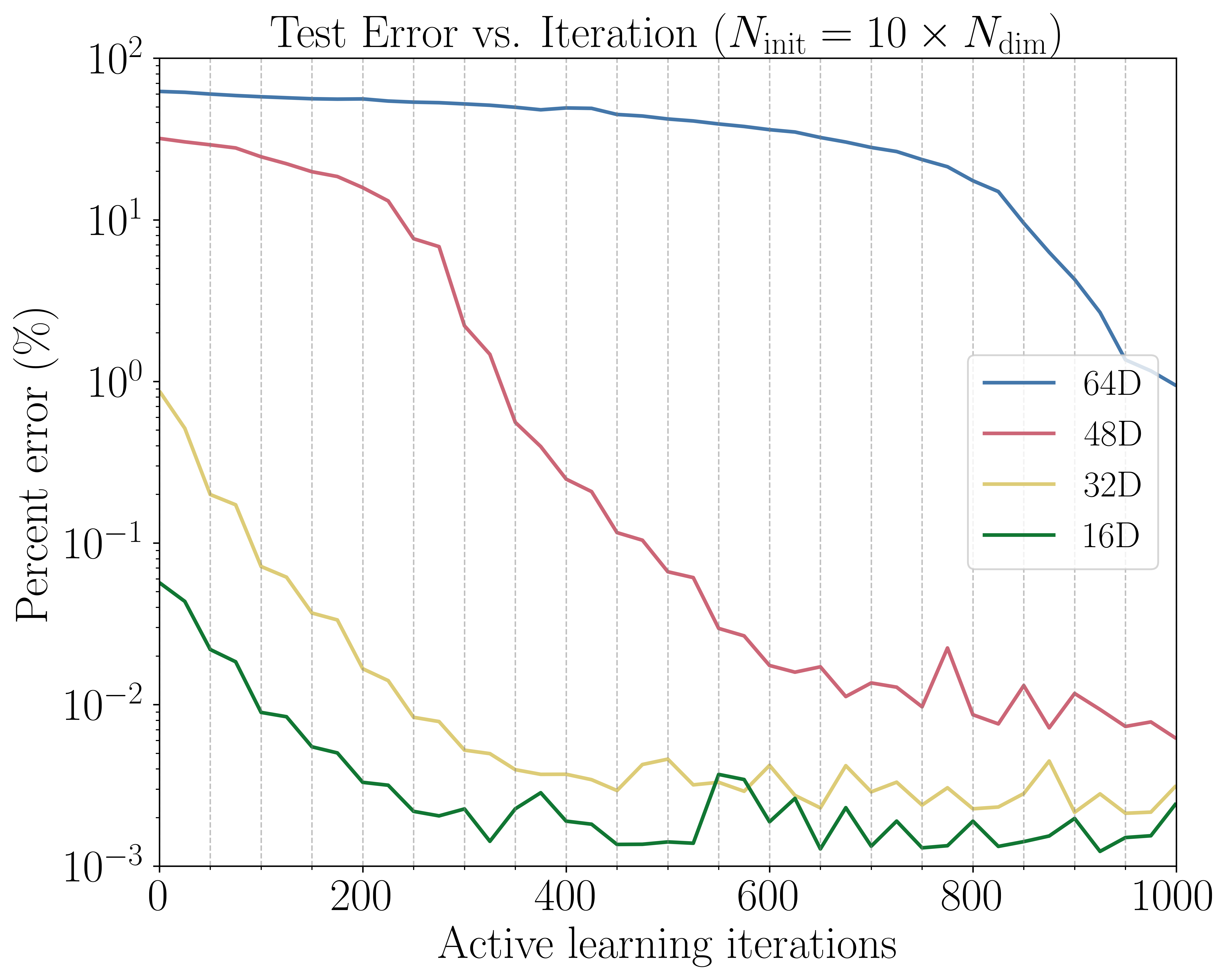}
    \caption{Surrogate model convergence as a function of active learning iterations (x-axis) and number of dimensions (colored lines). For each dimension, we train a surrogate model with $10\times N_{\rm dim}$ initial training samples and compute how well the model converges in terms of the percent error (y-axis) computed using test samples held out from the training sample. The vertical dashed lines show the iterations in which we re-optimized the hyperparameters (every 50 iterations).} 
    \label{fig:error_vs_iteration_10x}
\end{figure}

Next, we estimate the number of training samples required for the surrogate model to converge when trained on Gaussian distributions with 16, 32, 48, and 64 dimensions. Figure \ref{fig:error_vs_iteration_10x} shows the test error measured as the average percent error. Here $N_{\rm init}$ refers to the number of initial training samples, $N_{\rm active}$ refers to the number of active learning samples, and $N_{\rm dim}$ refers to the number of dimensions of the problem.
For each dimension shown with different colored lines, we train the GP using $10 \times N_{\rm dim}$ initial samples and run for an additional 1000 active learning iterations. We train the GP using an exponential squared kernel, perform active learning with the BAPE algorithm, and optimize the hyperparameters every 50 iterations using a 16-fold cross validation. 

\begin{figure*}
    \includegraphics[width=\linewidth]{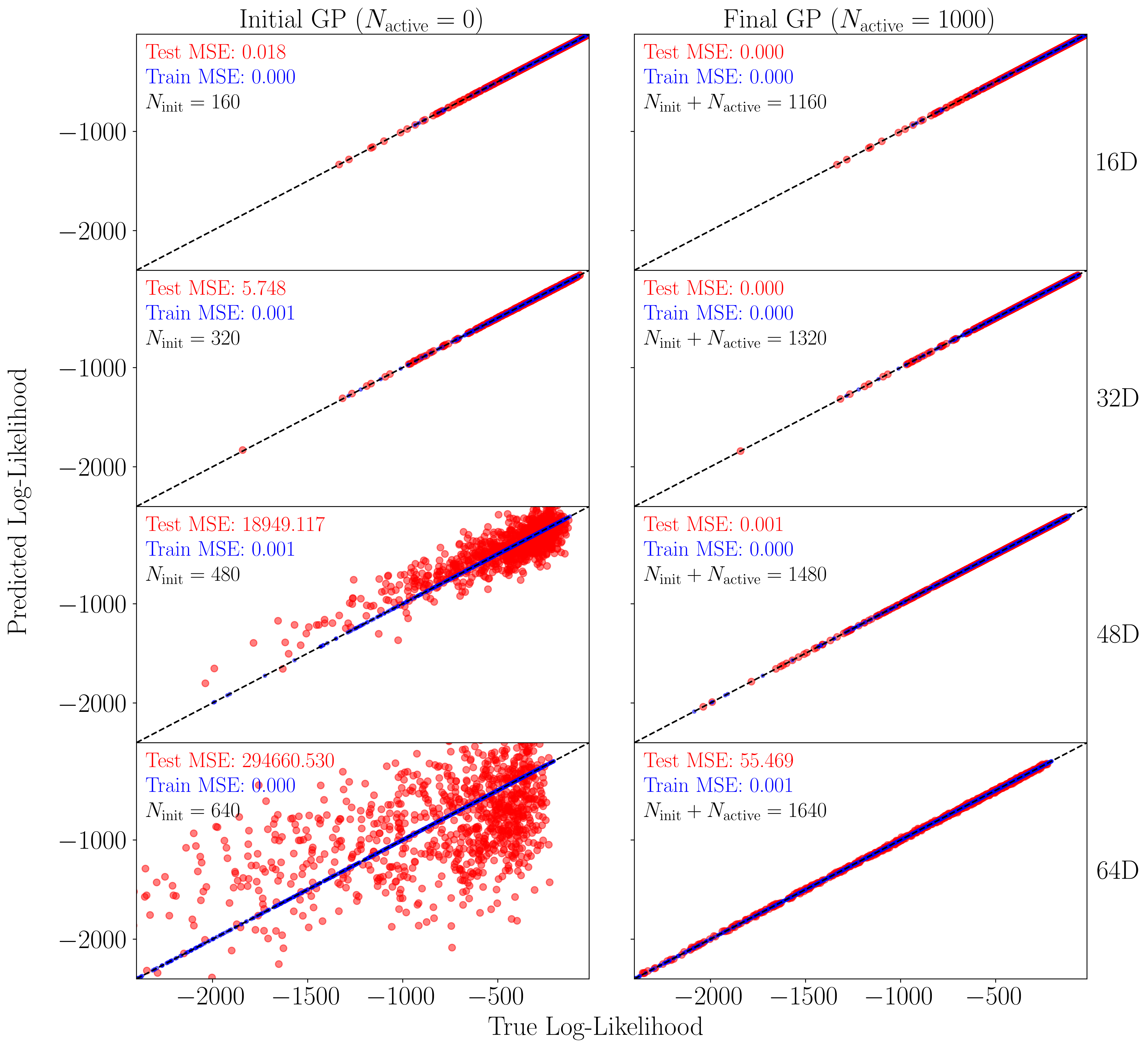}
    \caption{True log-likelihood (x-axis) vs. the predicted log-likelihood (y-axis), demonstrating the accuracy of the surrogate model for the initial GP fit (left column) and the final GP fit (right column). The initial GP is trained on $N_{\rm init}=10\times N_{\rm dim}$ samples, and the final GP is trained on $N_{\rm init} + N_{\rm active} = 10\times N_{\rm dim} + 1000$ samples. The rows show the initial and final GP fits for 8, 16, 32, 48, and 64 dimensional Gaussians, taken from the same runs as Figure \ref{fig:error_vs_iteration_10x}. The blue points show the predictions and mean squared error training sample, and the red points show that of the test sample.} 
    \label{fig:scatterplot_10x}
\end{figure*}

Figure \ref{fig:scatterplot_10x} shows the true log-likelihood vs. the predicted log-likelihood of the surrogate model. The left column shows the initial GP fit compared to the final GP fit after the 1000 active learning iterations. For $\leq32$ dimensions, the GP converges to less than 1\% error, given only the initial training samples ($N_{\rm init}=10\times N_{\rm dim}$). For 48 and 64 dimensions, the fit is initially poor, but converges to $<1\%$ error within 400 and 1000 iterations respectively. In all cases shown in Figure \ref{fig:scatterplot_10x}, the GP fits well to all of the training sample points, but fits poorly to the test sample in some cases (e.g., the initial fit for 48D and 64D). These cases with low training error but poor test error mean that the GP is not generalizing well to points that it wasn't trained on, showing that the GP can be prone to overfitting without enough training data. 

\clearpage

\subsection{Training data scaling} \label{subsec:training_data_scaling}

As discussed in Section~\ref{subsec:training_data}, scaling the input training data is an important preprocessing step that affects how the GP hyperparameters---particularly the kernel length scales---are optimized. When inputs are scaled to a common range (e.g., $[0,1]$ via a min-max scaler), the length scales become easier to interpret and constrain: values of order $\sim 0.1$--$1$ correspond to correlations spanning a meaningful fraction of the parameter domain. Without scaling, the length scale bounds must be matched to the raw coordinate ranges of each parameter, which vary between problems.

To demonstrate the practical effect of length scale bounds on surrogate model quality, we train a GP on a 4-dimensional correlated Gaussian likelihood:
\begin{equation}
    f_6(\mathbf{x}) = \mathcal{N}\left(
    \mathbf{0},\;
    \begin{bmatrix}
    1.00 & 0.60 & 0.30 & -0.20 \\
    0.60 & 1.00 & 0.10 & 0.40 \\
    0.30 & 0.10 & 1.00 & 0.50 \\
    -0.20 & 0.40 & 0.50 & 1.00
    \end{bmatrix}
    \right),
    \label{eqn:4d_gaussian}
\end{equation}
with each parameter defined over the domain $\theta_i \in [-4, 4]$. Both configurations use $N_{\rm init}=40$ initial Latin hypercube samples, $N_{\rm active}=60$ BAPE active learning iterations, the exponential squared kernel, min-max input scaling to $[0,1]$, L-BFGS-B hyperparameter optimization, and cross-validation re-optimized every 20 iterations. The configurations differ only in the allowed range of the kernel length scale hyperparameters, as summarized in Table~\ref{tab:lengthscale_configs}.

\begin{table}[!ht]
    \centering
    \caption{Hyperparameter configurations for the 4D Gaussian length scale comparison. Both configurations share the same kernel, optimizer, input scaling, and training procedure. The only difference is the allowed range of the GP length scale hyperparameters $\ell$.}
    \begin{tabular}{lcc}
        \hline
        \hline
        Setting & Config A & Config B \\
        \hline
        Kernel & ExpSquared & ExpSquared \\
        $\theta$ scaling & MinMax $[0,1]$ & MinMax $[0,1]$ \\
        HP opt.\ method & L-BFGS-B & L-BFGS-B \\
        $\log_{10}(\ell)$ range & $[-1, 1]$ & $[-4, -2]$ \\
        $\log_{10}(A)$ range & $[3, 5]$ & $[3, 5]$ \\
        White noise ($\log \sigma_n$) & $-12$ (fixed) & $-12$ (fixed) \\
        Fit amplitude & True & True \\
        Fit mean & True & True \\
        \hline
    \end{tabular}
    \label{tab:lengthscale_configs}
\end{table}

\alabi\ includes a \code{plot\_gp\_predictions\_1D} diagnostic that visualizes the GP surrogate by sweeping each input parameter individually while holding the remaining parameters fixed at a reference point (typically the surrogate maximum found via \code{find\_max\_surrogate}). For each dimension, the plot shows the GP predictive mean, the $\pm 2\sigma$ uncertainty band, individual function draws sampled from the GP posterior, and the training data. This diagnostic is particularly useful in problems with more than two dimensions, where 2D contour plots can only show pairwise projections.

\begin{figure*}[ht!]
    \centering
    \includegraphics[width=\linewidth]{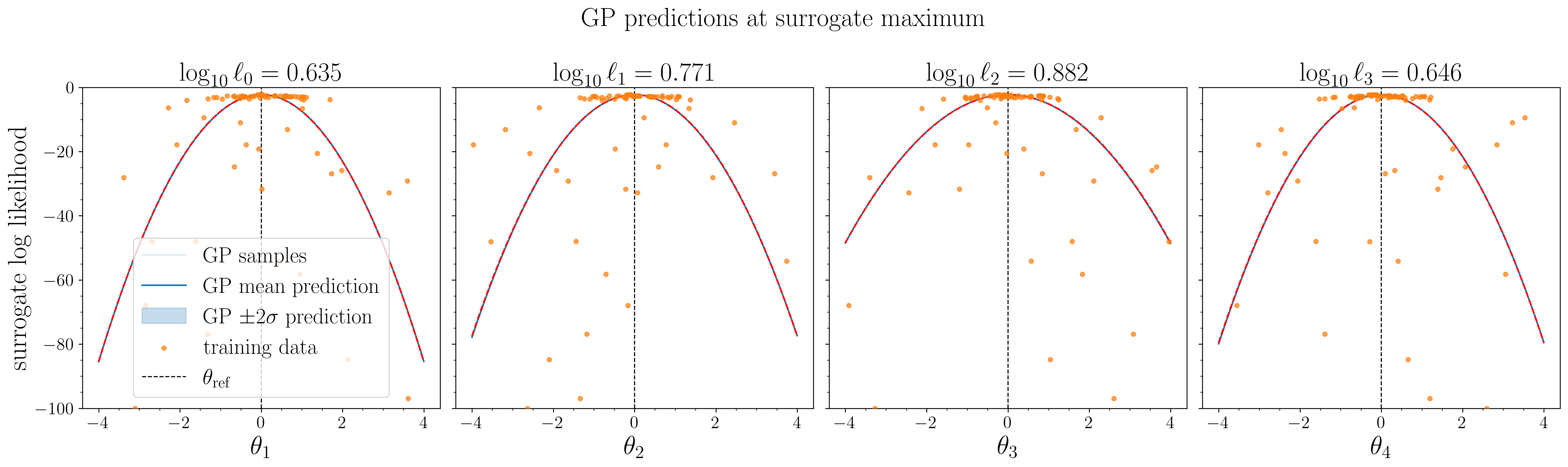}
    \includegraphics[width=\linewidth]{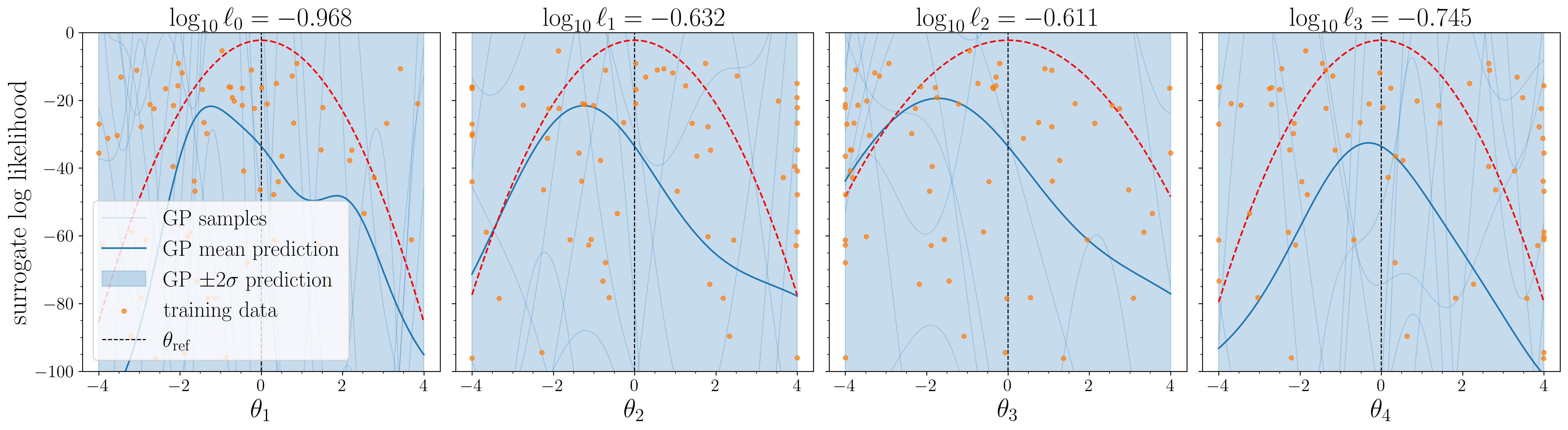}
    \caption{1D GP prediction diagnostic for Configuration~A (top panel), with well-chosen length scale bounds ($\log_{10}(\ell) \in [-1, 1]$) v.s. Configuration~B (bottom panel), with poorly chosen length scale bounds ($\log_{10}(\ell) \in [-4, -2]$). Each panel sweeps one parameter through its full range while holding the other three fixed at the surrogate maximum. The blue line shows the GP predictive mean, the shaded blue region shows the $\pm 2\sigma$ uncertainty, the light blue lines show individual function draws from the GP posterior, the orange points show the training data, and the dashed red line shows the true log-likelihood. For Configuration~A, the optimized length scales ($\log_{10}\ell \approx 0.2$--$0.5$) produce smooth, well-constrained predictions. In Configuration~B, the GP posterior draws are highly oscillatory, the uncertainty bands are much wider, and the mean prediction deviates from the true log-likelihood (dashed red line). This overfitting behavior results from length scales that are far too short for the smoothness of the underlying function.  \\}
    \label{fig:gp_1d_scales}
\end{figure*}

Figure~\ref{fig:gp_1d_scales} shows the 1D diagnostic for Configuration~A, where the length scale bounds are set to $\log_{10}(\ell) \in [-1, 1]$. Because the inputs have been scaled to $[0,1]$, length scales of order $\sim 0.1$--$1$ are appropriate for capturing the broad, smooth structure of the Gaussian likelihood. The optimized length scales are $\log_{10}(\ell) \approx 0.2$--$0.5$ across the four dimensions. The GP mean closely tracks the true log-likelihood (dashed red line), the uncertainty bands are narrow near training data, and the individual GP draws are smooth and well-behaved.

Figure~\ref{fig:gp_1d_scales} shows the same diagnostic for Configuration~B, where the length scale bounds are restricted to $\log_{10}(\ell) \in [-4, -2]$, corresponding to length scales of $10^{-4}$--$10^{-2}$. These very short length scales force the GP to treat each training point as nearly independent, producing a highly oscillatory interpolation. The individual GP posterior draws vary wildly between training points, the uncertainty bands are substantially wider, and the GP mean deviates significantly from the true model. This is a clear signature of overfitting: the GP memorizes the training data but fails to learn the underlying smooth structure of the likelihood surface. In practice, a surrogate model exhibiting this behavior would produce unreliable posterior estimates when sampled with MCMC.

This comparison illustrates several practical guidelines for diagnosing and avoiding overfitting due to length scale misconfiguration. Firstly, it is usually good practice to scale inputs before setting length scale bounds. When inputs are scaled to $[0,1]$, length scales of order $\sim 0.1$--$1$ (i.e., $\log_{10}(\ell) \in [-1, 1]$) are a reasonable starting range for most smooth functions.
Secondly, a good visual diagnostic for checking length scale overfitting is to plot the 1D fits around a given point (e.g. the maximum likelihood estimate; Figure~\ref{fig:gp_1d_scales}). If the GP posterior draws appear highly oscillatory (``wiggly'') or the uncertainty bands are unexpectedly wide, the length scales are likely too short. 
Finally, it is good to check the optimized length scale values and check if any of the values are converging towards the lower end of the bounds. In this case, the user may need to adjust the length scale regularization penalty (Equation~\ref{eqn:scale_reg}) to a higher mean and sigma value.

\subsection{Hyperparameter optimization choices} \label{subsec:hp_choices}

The quality of the GP surrogate model depends critically on the choice of hyperparameters used during training. Key choices include the kernel function, data scaling strategy, and optimization settings. Rather than manually tuning these, \alabi\ supports systematic testing of multiple configurations to identify the best settings for a given problem. In this section, we demonstrate this approach using the eggbox benchmark function (Example~\ref{subsubsec:eggbox}).

We define a base GP configuration and then vary three settings: the kernel function (exponential squared, Matern-3/2, and Matern-5/2), the input scaler (no scaling, min-max, and standard scaler), and the output scaler (no scaling, min-max, and standard scaler). This produces $3 \times 3 \times 3 = 27$ combinations. Table~\ref{tab:hp_base_config} summarizes the base configuration shared by all combinations, and Table~\ref{tab:hp_grid} shows the variable settings tested.

\begin{table}[!ht]
    \centering
    \caption{Base GP configuration shared across all hyperparameter (HP) combinations. These settings are held fixed while the kernel, input scaler, and output scaler are varied.}
    \begin{tabular}{lc}
        \hline
        \hline
        Setting & Value \\
        \hline
        Fit amplitude & True \\
        Fit mean & True \\
        Fit white noise & False \\
        White noise ($\log \sigma_n$) & $-12$ \\
        HP opt.\ method & L-BFGS-B \\
        HP selection & 8-fold cross-validation \\
        $\log_{10}(A)$ range & $[-1, 1]$ \\
        $\log_{10}(\ell)$ range & $[-2, 2]$ \\
        $N_{\rm init}$ & 50 \\
        Initial sampler & Latin hypercube \\
        \hline
    \end{tabular}
    \label{tab:hp_base_config}
\end{table}

\begin{table}[!ht]
    \centering
    \caption{Variable hyperparameter settings tested in the grid search. Each combination of kernel, input scaler ($\theta$), and output scaler ($y$) is evaluated by fitting a GP to the initial training sample and computing the test set MSE on 1000 held-out points.}
    \begin{tabular}{lc}
        \hline
        \hline
        Setting & Options tested \\
        \hline
        Kernel & ExpSquared, Matern-3/2, Matern-5/2 \\
        $\theta$ scaler & None, MinMax, Standard \\
        $y$ scaler & None, MinMax, Standard \\
        \hline
    \end{tabular}
    \label{tab:hp_grid}
\end{table}

For each of the 27 configurations, we fit the GP to the initial training sample of $N_{\rm init}=50$ points and evaluate performance using the mean squared error (MSE) on a held-out test set of 1000 points. Table~\ref{tab:hp_top5} shows the five best-performing configurations ranked by test MSE.

\begin{figure}[ht!]
    \centering
    \includegraphics[width=\linewidth]{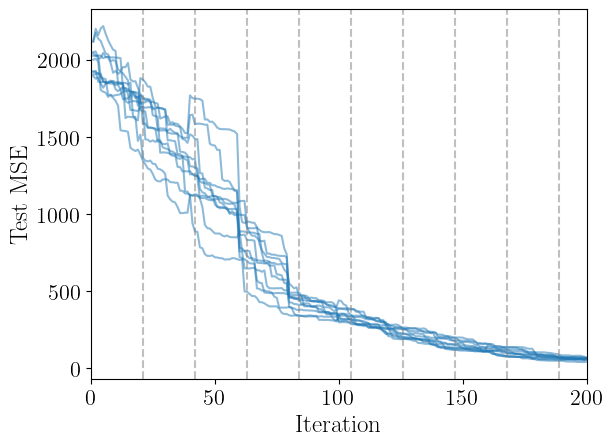}
    \caption{Test set MSE over 10 independent trials using the same hyperparameter configuration (Config~7 in Table \ref{tab:hp_top5}). Each blue line represents a different random seed. The spread illustrates the stochastic variance in the active learning process, while the overall convergence trend is consistent.}
    \label{fig:hp_variance}
\end{figure}

Using the best configuration (Config~7), we then run 200 active learning iterations with the BAPE algorithm, re-optimizing the GP hyperparameters every 20 iterations. Figure~\ref{fig:hp_variance} shows the test MSE as a function of active learning iteration. 
To assess the effect of random variance in the training process, we repeat the same configuration 10 times with different random seeds. Figure~\ref{fig:hp_variance} shows the test MSE trajectories for all 10 trials. The test MSE generally decreases steadily from $\sim 2000$ to $\sim 50$ over 200 iterations with some random variations between the 10 trials generated by different random seeds. While there is some spread in the convergence paths---particularly during the first $\sim 100$ iterations where different active learning samples explore different regions of parameter space---all trials converge to similarly low test MSE values by iteration 200.

\begin{figure}[ht!]
    \centering
    \includegraphics[width=\linewidth]{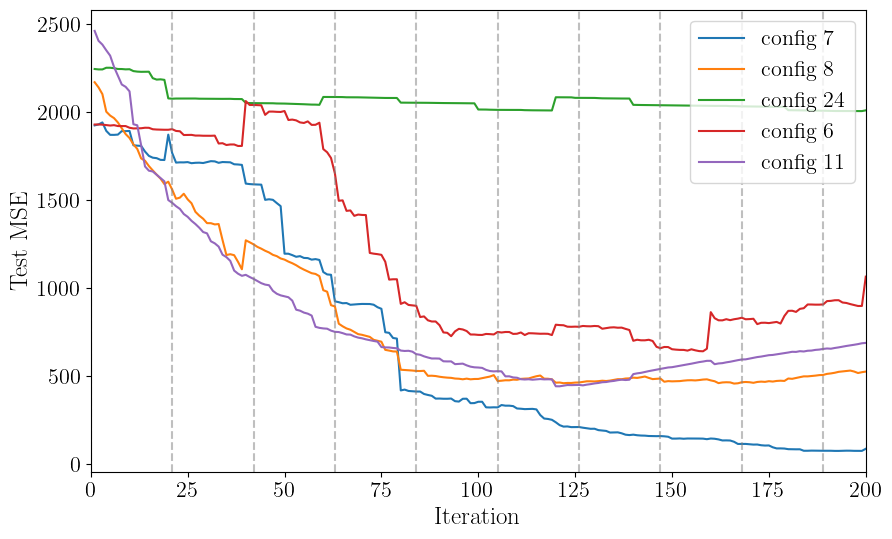}
    \caption{Comparison of active learning convergence for the top 5 hyperparameter configurations from Table~\ref{tab:hp_top5} on the eggbox benchmark. Each colored line corresponds to a different configuration. The best initial fit (Config~7, blue) also achieves the best final performance, but the relative ranking of other configurations changes during training.}
    \label{fig:hp_config_comparison}
\end{figure}

Finally, we compare the active learning performance of the top 5 initial configurations from Table~\ref{tab:hp_top5}. Figure~\ref{fig:hp_config_comparison} shows the test MSE trajectories for each. The configuration with the lowest initial test MSE (Config~7, exponential squared kernel with standard input scaling and min-max output scaling) also achieves the lowest final MSE after 200 active learning iterations. However, the ranking among configurations can change during active learning: Config~11 (Matern-3/2 with no input scaling) starts with a higher initial MSE but converges faster than Configs~24 and~6. Notably, Config~24 (Matern-5/2 with standard input scaling) shows minimal improvement during active learning, indicating that the initial hyperparameter fit quality does not always predict active learning performance. In practice, if the test error plateaus or fails to converge, this may indicate that a different hyperparameter configuration should be tried.

\begin{table}[!ht]
    \centering
    \caption{Top 5 hyperparameter configurations ranked by initial test set MSE for the eggbox benchmark. All configurations use the base settings from Table~\ref{tab:hp_base_config}. The best configuration uses the exponential squared kernel with standard input scaling and min-max output scaling.}
    \begin{tabular}{clllr}
        \hline
        \hline
        Config & Kernel & $\theta$ scaler & $y$ scaler & Test MSE \\
        \hline
        7 & ExpSquared & Standard & MinMax & 1993.4 \\
        8 & ExpSquared & Standard & Standard & 1998.5 \\
        24 & Matern-5/2 & Standard & None & 2029.1 \\
        6 & ExpSquared & Standard & None & 2066.4 \\
        11 & Matern-3/2 & None & Standard & 2067.5 \\
        \hline
    \end{tabular}
    \label{tab:hp_top5}
\end{table}

In conclusion, there is no definitive choice of hyperparameter configurations that will work for every problem, and it often takes a bit of experimenting to find which configuration is suited for your problem. GP optimization and test error performance are inherently variable (Figure~\ref{fig:hp_variance}), and do not necessarily converge monotonically (sometimes the test error will increase for some iterations before improving). As we show in this section, testing a grid of different configurations is a principled and practical way of selecting training configuration parameters without adding a lot of computational overhead (i.e. it requires recomputing the GP multiple times, but uses the existing training sample and doesn't require computing any new forward model evaluations). \\


\section{Discussion} \label{sec:discussion}

In this section, we discuss \alabi's performance in terms of accuracy and computational speed, as well as known limitations.
\alabi's accuracy at reproducing the true posterior depends on two key results: (\ref{subsec:convergence}) how well the GP surrogate model has converged to the true function, and (\ref{subsec:mcmc_discussion}) how well the MCMC sampler is able to accurately sample the GP surrogate model.

\subsection{How many training points should I use to train the surrogate model?} \label{subsec:convergence}

The amount of training points needed generally depends on the complexity and dimensionality of the problem. As a rule of thumb, a good starting place is to run $\sim10\times N_{\rm dim}$ initial training samples to train the model. For well ``behaved posteriors'' (posteriors with unimodal structure contained within a reasonably constrained prior space), we find that initial samples of $\sim10\times N_{\rm dim}$ are adequate, including for a relatively high number of dimensions. Depending on the error after the initial samples, additional active learning samples can be run as needed. \alabi's training can also be parallelized over multiple cores, allowing for the user to parallelize computing the initial training/test samples, as well as running multiple active learning chains at a time.

For more complex posteriors (those with multiple modes or degeneracies features), the user will likely need to train more/longer active learning chains to capture the high-likelihood regions. It is difficult to say a priori the exact number of samples that will be needed for a given problem, but a general strategy would be to run additional batches of $\sim10\times N_{\rm dim}$ active learning samples (either sequentially or divided over multiple processes) and repeat until the surrogate model has converged to the desired accuracy based on the chosen test error metric (Section \ref{subsubsec:test_error}). 

\subsection{How do I tell if the surrogate model has converged?}

Computing the prediction error (e.g. mean squared error, mean absolute error) is a straightforward and practical way to test how accurately the surrogate model is performing. Ideally, the user would compute at least $\sim10 \times N_{\rm dim}$ test samples, distributed across the prior space of interest to compute the error. The ``cutoff'' test error value depends on the users problem and the desired level of accuracy. For Figures \ref{fig:error_vs_iteration_10x}, we show the average percent error (\ref{eqn:percent_error}) so that the different examples can be compared relatively to one another.


\begin{figure}[h!]
    \includegraphics[width=\linewidth]{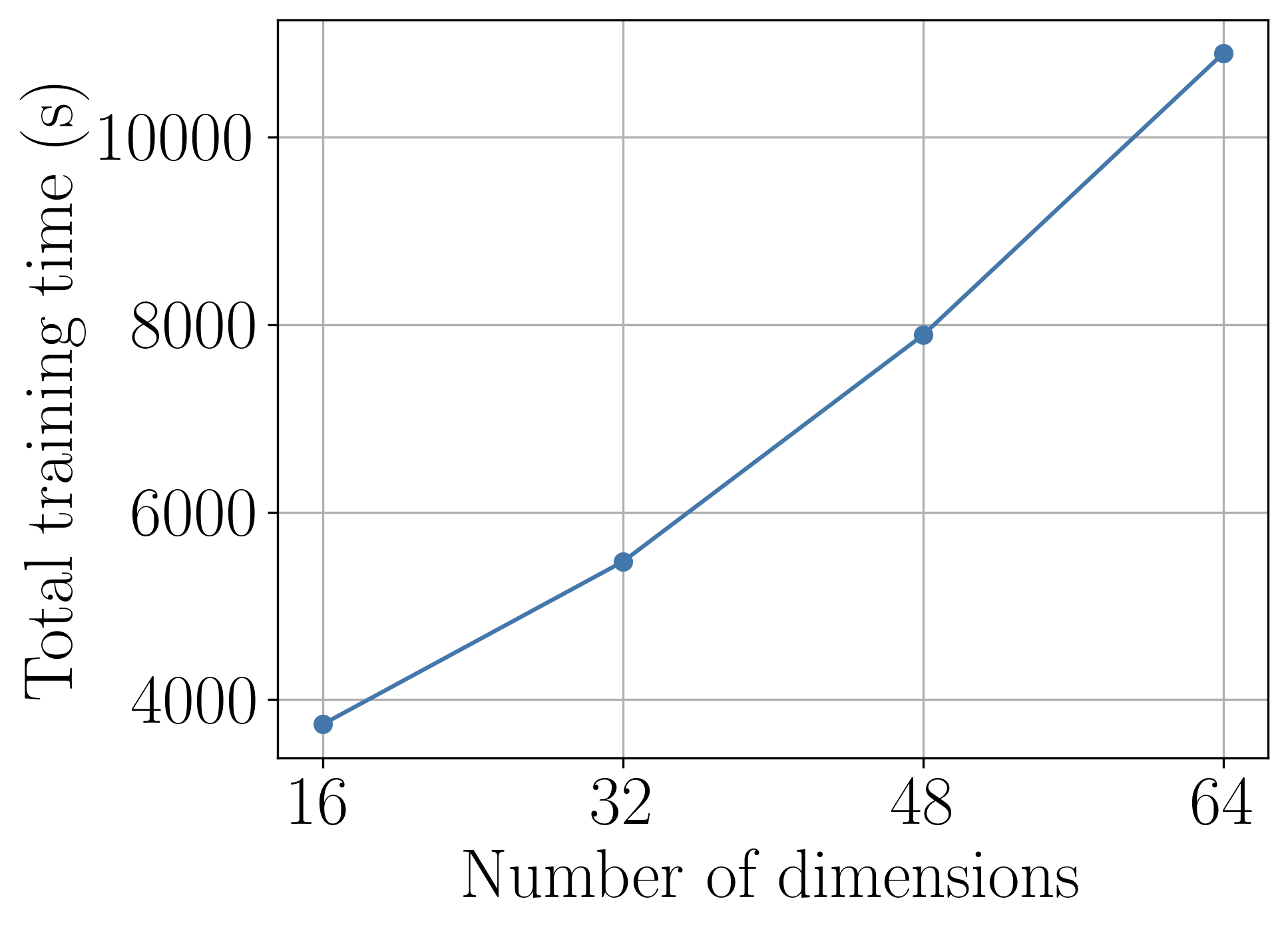}
    \caption{Cumulative training time as a function of active learning iterations for different numbers of dimensions.}
    \label{fig:cumm_train_time}
\end{figure}

\begin{figure}[h!]
\begin{center}
    \includegraphics[width=\linewidth]{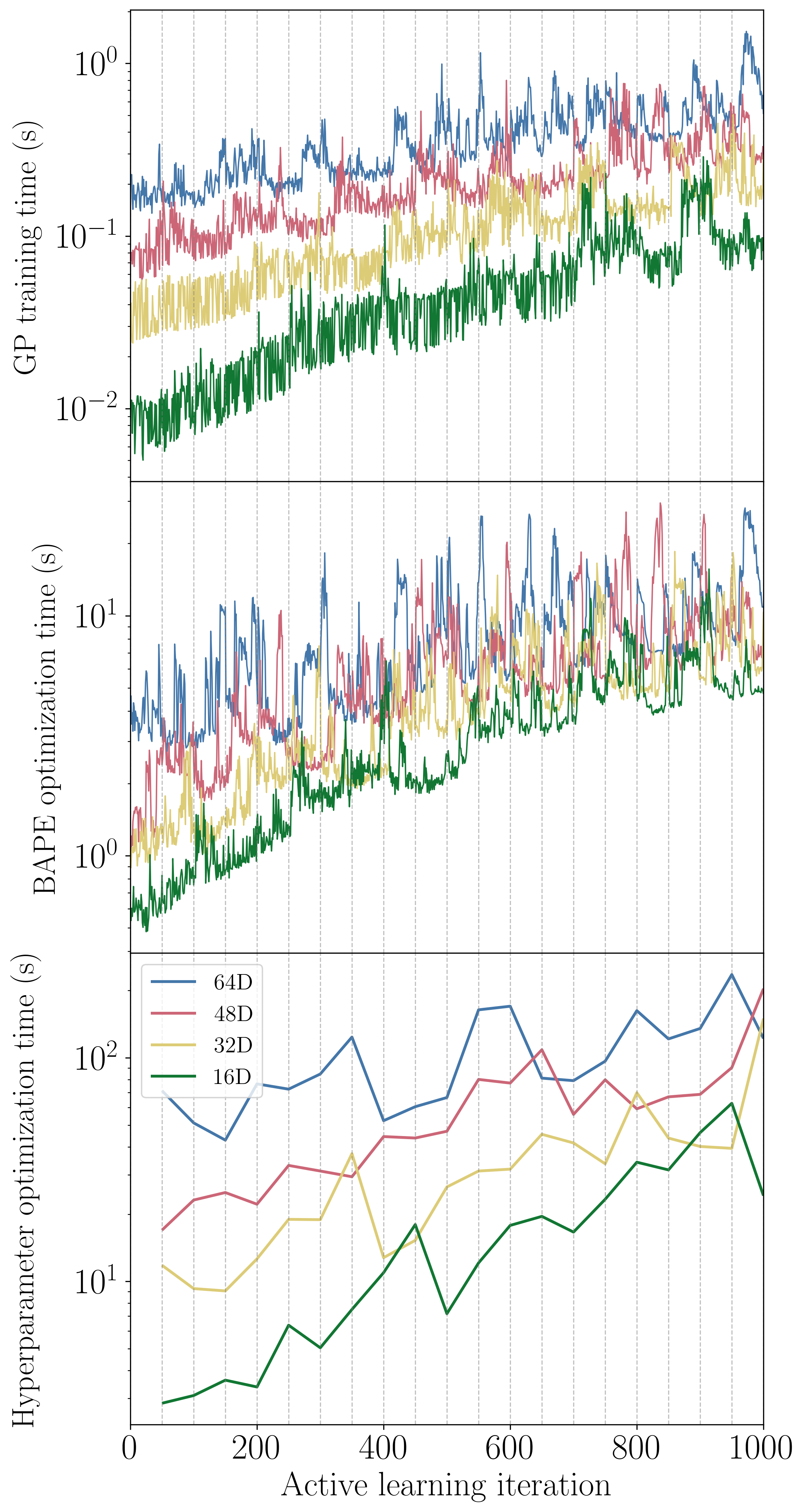}
    \vspace{10pt}
    \caption{The computation time for training \alabi\ as a function of the number of active learning iterations and number of dimensions. Each run (16, 32, 48 and 64D) is started with an initial training sample of $10\times N_{\rm dim}$ points, and run for 1000 active learning iterations, each iteration adding one additional training point. The top panel shows the time  to train the GP on the new data points each iteration, the middle panel shows the time to optimize the BAPE active learning function each iteration, and the bottom panel shows the time to reoptimize the GP hyperparameters every 50 iterations. The primary computational overhead for training the GP surrogate is the training points used, and the total computation time scales exponentially with the number of training points.}
    \label{fig:time_scaling}
\end{center}
\end{figure}

\begin{figure}[h!]
    \centering
    \includegraphics[width=\linewidth]{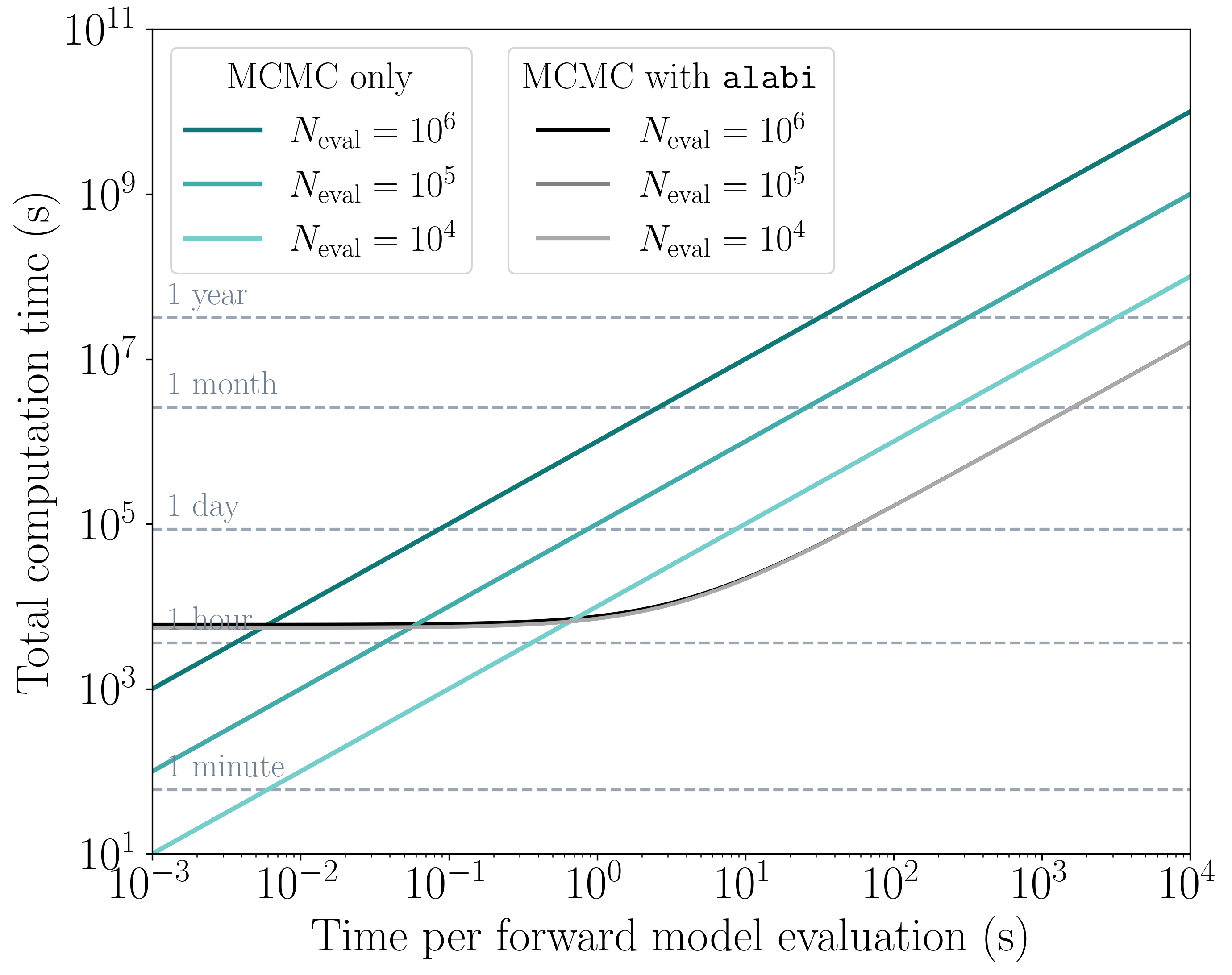}
    \caption{Total MCMC computation time as a function of forward model evaluation time. The teal lines show the total MCMC time when running the sampler with the true model for varying numbers of samples ($N_{\rm eval} = 10^4, 10^5, 10^6$), while the gray lines show the total MCMC time when running the sampler with the \alabi surrogate model. For forward models taking $\lesssim 1$s to evaluate, the \alabi training overhead exceeds the exceeds the computational savings for computing MCMC samples, however for $\gtrsim 1$s, MCMC with \alabi performs a factor of $10-1000\times$ faster, depending on the number of posterior samples computed.}
    \label{fig:computational_scaling}
\end{figure}

\subsection{Which posterior sampler should I use for my problem?} \label{subsec:mcmc_discussion}

Assuming that the surrogate model can converge to the true posterior, \alabi can only perform as well as the underlying sampler that the user chooses. This sampling performance can depend on the posterior structure and number of dimensions.

For any sort of problem with ``complex'' features, such as multiple modes or degeneracies, nested sampling (through \dynesty, \multinest, or \ultranest) tends to be much more reliable than affine-invariant sampling (such as \emcee). For unimodal problems however, \emcee offers more flexibility for easily defining custom prior functions, unlike nested sampling, which requires defining a prior transform function that maps points from a unit cube to the desired function. \emcee is also simpler to run than nested sampling, requiring no free parameters to tune, whereas nested sampling has many.

Posterior sampling for high dimensional problems is also an open area of research \citep{jones_hobert_2001}. Discussing the limitations of sampling methods is beyond the scope of this paper, however, we refer the reader to \cite{Dittmann_2024} for a discussion on hyperparameter tuning for nested sampling methods, as well as \cite{rajaratnam2015} for a discussion on convergence for MCMC methods in high dimensions.  

\subsection{When is it worth training a surrogate model for my likelihood?} \label{subsec:cpu_time}

\alabi\ can significantly reduce the number of forward model evaluations by orders of magnitude. However, the process of training \alabi\ also introduces some computational overhead, and it isn't fully advantageous to use \alabi\ unless one's forward model is computationally slow. Figure \ref{fig:time_scaling} breaks down how much computational overhead goes into each step of \alabi\ for the GP training time, objective function optimization, and hyperparameter optimization.

Figure \ref{fig:time_scaling} shows how \alabi's training time scales as a function of training iterations and number of dimensions. The top panel shows the scaling of the GP training time, which is the computation time required to compute the predictive mean and variance (Eqn. \ref{eqn:gp_mean} and \ref{eqn:gp_var}) by inverting the covariance matrix. The computational cost of inverting the covariance scales exponentially with the number of data points as $\mathcal{O}(n \log^2 n)$ using the \code{george} implementation of the GP \citep{ambikasaran_fast_2015}. 

For training the GP, the computation time scales exponentially with the number of training samples, but for $\leq32$ dimensions, it is not significantly more costly to run for a larger number of dimensions for a stationary kernel. 
The second panel of Figure \ref{fig:time_scaling} shows how long it takes to optimize the active learning function, used to select the next training point. In this example, we optimize the BAPE active learning function (\ref{eqn:bape}) during each iteration. The number of dimensions of the acquisition function is equal to the number of dimensions of the likelihood. 
Lastly, the bottom panel of Figure \ref{fig:time_scaling} shows the amount of time it takes to reoptimize the hyperparameters of the GP, which is performed every 50 active learning iterations. The number of hyperparameters is equal to the number of dimensions (one length scale parameter per dimension), plus two parameters (for the mean and amplitude parameters).

Figure \ref{fig:computational_scaling} shows how the cumulative training time translates to the total MCMC sampling time (y-axis) as a function of forward model evaluation time (x-axis) for varying numbers of MCMC samples--i.e., the number of times the MCMC evaluates the model (teal showing for the true model, and gray showing for the surrogate model). For forward model evaluations $\gtrsim 1$s, MCMC with the \alabi performs a factor of $10-1000\times$ faster than MCMC with the true model, depending on the number of posterior samples computed.
For \alabi, the main computational overhead for training a GP surrogate model is mainly dependent on the number of training samples. Models with more dimensions take a larger number of training samples to converge, hence, the cumulative training time for higher dimensional problems is longer.

\subsection{What hyperparameter settings should I use for my problem?} \label{subsec:hp_guidance}

In practice, selecting GP hyperparameter settings (including the kernel function, data scaling, length scale bounds, and optimization method) is one of the most challenging aspects of training a surrogate model. A poor choice of settings can lead to overfitting, slow convergence, or numerically unstable fits, and the symptoms are not always obvious from the training loss alone. Sections~\ref{subsec:training_data_scaling} and~\ref{subsec:hp_choices} demonstrate two complementary strategies for addressing this challenge: visual diagnostics of GP fit quality, and systematic grid search over candidate configurations. Together, these provide a principled workflow for selecting hyperparameters.

We recommend the following approach: first, start with a grid search on the initial training sample. As demonstrated in Section~\ref{subsec:hp_choices}, fit the GP using a grid of candidate configurations (varying the kernel, input scaler, and output scaler) and rank them by test set MSE on held-out points (Table~\ref{tab:hp_top5}). This step is inexpensive because it only requires fitting the GP to the initial training sample (no additional forward model evaluations), and it quickly narrows the search space to a small number of promising settings. The grid search can be performed after the initial sampling phase (via \code{init\_samples}) and before starting active learning.

For the top-ranked configurations, plot the 1D GP predictions around the maximum likelihood point (Section~\ref{subsec:training_data_scaling}) to visually inspect the GP fit (Figure~\ref{fig:gp_1d_scales}). In general, you want to check that: (a) the GP mean tracks the training data without excessive oscillation, (b) the uncertainty bands are reasonably narrow near training points, (c) the individual GP posterior draws are smooth and consistent, and (d) the optimized length scale values are not pinned at the boundary of their allowed range. If the GP draws appear overly ``wiggly'' or the uncertainty is unexpectedly large, the length scale bounds or penalty term (\ref{eqn:scale_reg}) likely need to be adjusted.

From there, run active learning in batches and re-evaluate. Rather than committing to a full active learning run with a single configuration, run a moderate number of iterations (e.g., 50--100) and then re-check the test MSE and 1D diagnostic. As shown in Figure~\ref{fig:hp_config_comparison}, the relative performance of configurations can change during active learning--a configuration that ranks well on the initial fit does not always converge the fastest. If the test error plateaus or increases, this is a signal to try a different configuration before investing further computational effort. This iterative approach prevents the user from wasting time on a setting that will not converge.

This workflow is designed to be practical: the grid search and diagnostic steps add minimal overhead relative to the cost of training the surrogate model, and they can save substantial time by identifying poor configurations early.

\subsection{Parallelization}

There are several processes that can be parallelized in \alabi's workflow: 
\begin{enumerate}
    \item \textbf{Initial training:} Computing the initial training and test samples can be done in parallel.
    \item \textbf{Optimizing hyperparameters:} for the marginal likelihood approach (\ref{subsubsec:ml_approach}) \alabi\ will run multiple optimizations in parallel, or for the cross-validation approach (\ref{subsubsec:cv_approach}) it will parallelize computing the cross-validation score over each candidate hyperparameter.
    \item \textbf{Active learning chains:} \alabi can run multiple active learning chains in parallel and combine all training samples at the end.
    \item \textbf{MCMC sampling:} \alabi can easily make use of the parallelized implementation for different samplers, allowing for parallelized sampling chains.
\end{enumerate}
So while steps $1-4$ must be run sequentially, the computations within each step can be distributed over multiple processors. For step 3, we find that parallelizing multiple chains while adding single points per iteration is more computationally efficient than running a single chain while adding multiple points per iteration. This efficiency arises for two reasons. First, because sequentially adding new points to the GP allows the GP to update the acquisition function each iteration, so that each new point informs the model. On the other hand, drawing multiple points at a time from the same acquisition function can cause multiple points to converge at the same location, producing redundant training data. Second, as noted in Section \ref{subsec:cpu_time}, the training time takes longer per iteration as the number of training samples increases. Therefore, by parallelizing over different chains, each chain remains relatively short (making computation time quicker) while adding unique information to the fit.

\subsection{Limitations \& Future Work}

This paper compares the performance of commonly used kernels, active learning algorithms and MCMC samplers appropriate for the task of performing Bayesian inference with active learning, but it is not an exhaustive test of all possible combinations of algorithms. However, \alabi\ is modularized so that new algorithms can be incorporated in the future, or custom defined by users. Further work would also implement more flexibility for different user-defined kernels. 

Currently \alabi's use of gradient optimization is performed by taking analytic gradients of functions, but an alternative GP backend that is built off of an auto-differentiable framework could allow for improved optimization speed and flexibility.  \\

\section{Conclusion} \label{sec:conclusion}

We have presented \alabi: an open-source Python framework for accelerating Bayesian inference with computationally expensive forward models. This tool presents a way forward for performing simulation based inference with complex simulations and datasets. This framework has been applied to a variety of computational astrophysics problems \citep[e.g.,][]{birky_improved_2021,birky_prospects_2025}, and can furthermore be used for inference with any black-box forward model in any domain of computational science.


We evaluated \alabi\ on a suite of benchmark problems spanning unimodal distributions, curved degeneracies, multimodal structure, and high dimensionality. Our main findings are:
(1) Accuracy: \alabi\ successfully trains surrogate models for problems with up to 64 dimensions (Section~\ref{subsubsec:nd_gaussian}), and can achieve $<1\%$ average percent error with $\sim10 \times N_{\rm dim}$ initial training points and up to 1000 active learning iterations. (2) Computational speedup: for forward models with evaluation times $\gtrsim 1$\,s, \alabi\ reduces the total number of model evaluations by orders of magnitude compared to direct MCMC sampling, yielding speedups of $10$--$1000\times$ (Section~\ref{subsec:cpu_time}). (3) Hyperparameter selection: We demonstrated a principled workflow for selecting GP hyperparameters (Sections~\ref{subsec:training_data_scaling}, \ref{subsec:hp_choices}, and~\ref{subsec:hp_guidance}) using a grid search and visual diagnostics test.

Each step of \alabi's workflow---initial sampling, hyperparameter optimization, active learning, and posterior sampling---can be parallelized across multiple processors, enabling the framework to scale to large computing clusters. \alabi\ is designed to be modular: new kernels, acquisition functions, and sampling backends can be incorporated as they become available. The package, along with comprehensive documentation and tutorials for all benchmark problems presented in this paper, is publicly available at \url{https://github.com/jbirky/alabi}. \\

\begin{acknowledgments} \label{sec:acknowledgments}
	The authors would like to acknowledge David Fleming (formerly UW), James Davenport (UW), Eric Agol (UW), Tyler Gordon (UCSC), and Samantha Gilbert-Janizek (UW) for various constructive discussions in the process of this project. 
	JB acknowledges funding support from NSF Graduate Research Fellowship grant number DGE-1762114 and a Scialog grant supported by the Heissing-Simmons Foundation.
\end{acknowledgments}

\software{
    \code{APPROXPOSTERIOR} \citep{fleming_approxposterior_2018}, \,
    \code{dynesty} \citep{speagle_dynesty_2020}, \,
    \code{emcee} \citep{foreman-mackey_emcee_2013}, \,
    \code{corner} \citep{foreman-mackey_cornerpy_2016}, \,
    \code{george} \citep{ambikasaran_fast_2015}, \,
    \code{matplotlib} \citep{hunter_matplotlib_2007}, \,
    \code{numpy} \citep{walt_numpy_2011}, \,
    \code{pymultinest} \citep{feroz_multinest_2009,Buchner2014}
    \code{scikit-optimize} \citep{head_scikit-optimizescikit-optimize_2020}, \,
    \code{scipy} \citep{jones_scipy_2001}, \,
    \code{ultranest} \citep{Buchner2021}. \\
}

The code for \alabi is available on Github: \href{https://github.com/jbirky/alabi}{https://github.com/jbirky/alabi}.

\clearpage
\bibliographystyle{aasjournal}

\end{document}